\input harvmac.tex
%  we don't use this tables macros
%\input tables.tex
\input amssym.tex

%\draftmode

%%%%%%%%%%%%%%%%%%%%%%%%%%%  REFERENCES  %%%%%%%%%%%%%%%%%%%%%%%%%%%%%

\lref\WittenJones{
E.~Witten, ``Quantum Field Theory And The Jones Polynomial,''
Commun.\ Math.\ Phys.\  {\bf 121}, 351 (1989).}

%\lref\HOMFLY{P.~Freyd, D.~Yetter, J.~Hoste, W.~Lickorish,
%K.~Millett, A.~Oceanu, ``A New Polynomial Invariant of Knots and Links,''
%Bull. Amer. Math. Soc. {\bf 12} (1985) 239.}

\lref\Turaev{
V.~G.~Turaev,
``The Yang-Baxter equation and invariants of links,''
Invent.\ Math.\ {\bf 92} 527--553 (1988).}

\lref\Wittencsstring{ E.~Witten,
``Chern-Simons gauge theory as a string theory,''
Prog.\ Math.\  {\bf 133} (1995) 637, hep-th/9207094.}

\lref\Wittengrass{E.~Witten,
``The Verlinde algebra and the cohomology of the Grassmannian,'' hep-th/9312104.}

\lref\GepnerGR{D.~Gepner,
``Fusion Rings And Geometry,''  Commun.\ Math.\ Phys.\  {\bf 141}, 381 (1991).}

\lref\OV{
H.~Ooguri, C.~Vafa, ``Knot Invariants and Topological Strings,''
Nucl.Phys. {\bf B577} (2000) 419.}

\lref\Khovanov{M.~ Khovanov,
``A categorification of the Jones polynomial,''
Duke Math. J. {\bf 101} (2000) 359, math.QA/9908171.}

\lref\Khovanovii{M.~ Khovanov,
``Categorifications of the colored Jones polynomial,''
J. Knot Theory Ramifications {\bf 14} (2005) 111, math.QA/0302060.}

\lref\Khovanoviii{M.~ Khovanov, ``$sl(3)$ link homology,''
Algebr. Geom. Topol. {\bf 4} (2004) 1045, math.QA/0304375.}

\lref\Khovanoviv{M.~ Khovanov,
``An invariant of tangle cobordisms,'' math.QA/0207264.}

\lref\Khovanovv{M.~Khovanov,
``Link homology and Frobenius extensions,''
math.QA/0411447.}

\lref\DBN{D.~Bar-Natan,
``On Khovanov's categorification of the Jones polynomial,''
Algebr. Geom. Topol. {\bf 2} (2002) 337, math.QA/0201043.}

\lref\DBNnews{D.~Bar-Natan, ``Some Khovanov-Rozansky Computations'' \semi
{http://www.math.toronto.edu/~drorbn/Misc/KhovanovRozansky/index.html}}

\lref\DBNi{D.~Bar-Natan, 
``Khovanov's Homology for Tangles and Cobordisms,''
math.GT/0410495.}

\lref\turner{P.~Turner,
``Calculating Bar-Natan's characteristic-two Khovanov homology,''
math.GT/0411225.}

\lref\Jacobsson{M.~Jacobsson,
``An invariant of link cobordisms from Khovanov's homology,''
Algebr. Geom. Topol. {\bf 4} (2004) 1211, math.GT/0206303.}

\lref\RKhovanov{M.~Khovanov, L.~Rozansky,
``Matrix factorizations and link homology,'' math.QA/0401268.}

\lref\Rfoam{L.~Rozansky,
``Topological A-models on seamed Riemann surfaces,'' hep-th/0305205.}

\lref\KRfoam{M.~Khovanov and L.~Rozansky,
``Topological Landau-Ginzburg models on a world-sheet foam,'' hep-th/0404189.}

\lref\GopakumarV{R.~Gopakumar and C.~Vafa,
``On the gauge theory/geometry correspondence,''
Adv.\ Theor.\ Math.\ Phys.\  {\bf 3} (1999) 1415, hep-th/9811131.}

\lref\GViii{R.~Gopakumar and C.~Vafa,
``M-theory and topological strings. I,II,''
hep-th/9809187; hep-th/9812127.}

\lref\SinhaV{S.~Sinha and C.~Vafa,
``SO and Sp Chern-Simons at large N,''
arXiv:hep-th/0012136.}

\lref\HoriBK{
K.~Hori, K.~Hosomichi, D.~C.~Page, R.~Rabadan and J.~Walcher,
``Non-perturbative orientifold transitions at the conifold,''
JHEP {\bf 0510}, 026 (2005), hep-th/0506234.}

\lref\KKV{S.~Katz, A.~Klemm and C.~Vafa,
``M-theory, topological strings and spinning black holes,''
Adv.\ Theor.\ Math.\ Phys.\  {\bf 3} (1999) 1445, hep-th/9910181.}

\lref\mirbook{``Mirror Symmetry'' (Clay Mathematics Monographs, V. 1),
K.~Hori et.al. ed, American Mathematical Society, 2003.}

\lref\HV{K.~Hori, C.~Vafa, ``Mirror Symmetry,'' hep-th/0002222;
K.~Hori, A.~Iqbal, C.~Vafa, ``D-Branes And Mirror Symmetry,'' hep-th/0005247.}

\lref\AKV{M.~Aganagic, A.~Klemm, C.~Vafa,
``Disk Instantons, Mirror Symmetry and the Duality Web,'' hep-th/0105045.}

\lref\AKMV{M.~Aganagic, A.~Klemm, M.~Marino and C.~Vafa,
``Matrix model as a mirror of Chern-Simons theory,''
JHEP {\bf 0402}, 010 (2004), hep-th/0211098.}

\lref\AAHV{B.~Acharya, M.~Aganagic, K.~Hori and C.~Vafa,
``Orientifolds, mirror symmetry and superpotentials,'' hep-th/0202208.}

\lref\LMV{J.~M.~F.~Labastida, M.~Marino and C.~Vafa,
``Knots, links and branes at large N,''
JHEP {\bf 0011}, 007 (2000), hep-th/0010102.}

\lref\LMtorus{J.~M.~F.~Labastida and M.~Marino,
``Polynomial invariants for torus knots and topological strings,''
Commun.\ Math.\ Phys.\  {\bf 217} (2001) 423, hep-th/0004196.}

\lref\LMqa{J.~M.~F.~Labastida and M.~Marino,
``A new point of view in the theory of knot and link invariants,''
math.qa/0104180.}

\lref\HSTa{S.~Hosono, M.-H.~Saito, A.~Takahashi,
``Holomorphic Anomaly Equation and BPS State Counting of Rational
Elliptic Surface,'' Adv.Theor.Math.Phys. {\bf 3} (1999) 177.}

\lref\HSTb{S.~Hosono, M.-H.~Saito, A.~Takahashi,
``Relative Lefschetz Action and BPS State Counting,''
Internat. Math. Res. Notices, (2001), No. 15, 783.}

\lref\Kprivate{M.~Khovanov, private communication.}

\lref\Taubes{
C.~Taubes, ``Lagrangians for the Gopakumar-Vafa conjecture,''
Adv. Theor. Math. Phys. {\bf 5} (2001) 139, math.DG/0201219.}

\lref\Wittenams{E.~Witten, ``Dynamics of Quantum Field Theory,''
{\it Quantum Fields and Strings: A Course for Mathematicians}
(P. Deligne, {\it et.al.} eds.), vol. 2, AMS Providence, RI, (1999) pp. 1313-1325.}

\lref\LVW{W.~Lerche, C.~Vafa and N.~P.~Warner,
``Chiral Rings In N=2 Superconformal Theories,''
Nucl.\ Phys.\ B {\bf 324}, 427 (1989).}

\lref\HMoore{J.~A.~Harvey and G.~W.~Moore,
``On the algebras of BPS states,''
Commun.\ Math.\ Phys.\  {\bf 197} (1998) 489, hep-th/9609017.}

\lref\IqbalV{T.~J.~Hollowood, A.~Iqbal and C.~Vafa,
``Matrix models, geometric engineering and elliptic genera,''
hep-th/0310272.}

\lref\Schwarz{A.~Schwarz,
``New topological invariants arising in the theory of quantized fields,''
Baku International Topological Conf., Abstracts (part II) (1987).}

\lref\SchwarzS{A.~Schwarz and I.~Shapiro,
``Some remarks on Gopakumar-Vafa invariants,'' hep-th/0412119.}

\lref\Aspinwallrev{ P.~S.~Aspinwall, ``D-branes on Calabi-Yau
manifolds,'' hep-th/0403166.}

\lref\MNOP{D.~Maulik, N.~Nekrasov, A.~Okounkov, R.~Pandharipande,
``Gromov-Witten theory and Donaldson-Thomas theory, I,'' math.AG/0312059.}

\lref\Katz{S.~Katz, ``Gromov-Witten, Gopakumar-Vafa, and Donaldson-Thomas
invariants of Calabi-Yau threefolds,'' math.ag/0408266.}

\lref\FultonH{W.~Fulton, J.~Harris,
``Representation Theory: A First Course,'' Springer-Verlag 1991.}

\lref\Kontsevich{M.~Kontsevich, unpublished.}

\lref\KapustinLi{A.~Kapustin and Y.~Li,
``D-branes in Landau-Ginzburg models and algebraic geometry,''
JHEP {\bf 0312} (2003) 005, hep-th/0210296.}

\lref\KapustinLii{A.~Kapustin and Y.~Li,
``Topological correlators in Landau-Ginzburg models with boundaries,''
Adv.\ Theor.\ Math.\ Phys.\  {\bf 7} (2004) 727, hep-th/0305136.}

\lref\KapustinLiii{A.~Kapustin and Y.~Li,
``D-branes in topological minimal models: The Landau-Ginzburg approach,''
JHEP {\bf 0407} (2004) 045, hep-th/0306001.}

\lref\Brunneriii{
I.~Brunner, M.~Herbst, W.~Lerche and J.~Walcher,
``Matrix factorizations and mirror symmetry: The cubic curve,''
hep-th/0408243.}

\lref\Brunner{I.~Brunner, M.~Herbst, W.~Lerche and B.~Scheuner,
``Landau-Ginzburg realization of open string TFT,'' hep-th/0305133.}

\lref\HLLii{M.~Herbst, C.~I.~Lazaroiu and W.~Lerche,
``D-brane effective action and tachyon condensation in topological minimal
models,'' hep-th/0405138.}

\lref\HLL{M.~Herbst, C.~I.~Lazaroiu and W.~Lerche,
``Superpotentials, A(infinity) relations and WDVV equations for open
topological strings,'' hep-th/0402110. }

\lref\calin{
  M.~Herbst and C.~I.~Lazaroiu,
  ``Localization and traces in open-closed topological Landau-Ginzburg
  models,''
  JHEP {\bf 0505}, 044 (2005)
  [arXiv:hep-th/0404184].}

\lref\LercheJW{W.~Lerche and J.~Walcher,
``Boundary rings and N = 2 coset models,''
Nucl.\ Phys.\ B {\bf 625} (2002) 97, hep-th/0011107.}

\lref\HoriJW{K.~Hori and J.~Walcher,
``F-term equations near Gepner points,''
JHEP {\bf 0501} (2005) 008, hep-th/0404196.}

\lref\Orlov{D.~Orlov, ``Triangulated Categories of Singularities
and D-Branes in Landau-Ginzburg Orbifold,''
Proc. Steklov Inst. Math. {\bf 246} (2004) 227, math.AG/0302304.}

\lref\Emanuelii{S.~K.~Ashok, E.~Dell'Aquila, D.~E.~Diaconescu and
B.~Florea, ``Obstructed D-branes in Landau-Ginzburg orbifolds,''
hep-th/0404167.}

\lref\Emanuel{ S.~K.~Ashok, E.~Dell'Aquila and D.~E.~Diaconescu,
``Fractional branes in Landau-Ginzburg orbifolds,''
hep-th/0401135.}

\lref\Rstability{
  J.~Walcher,
  ``Stability of Landau-Ginzburg branes,''
  J.\ Math.\ Phys.\  {\bf 46}, 082305 (2005)
  [arXiv:hep-th/0412274].
  %%CITATION = HEP-TH 0412274;%%
}

\lref\GovindarajanIM{
  S.~Govindarajan, H.~Jockers, W.~Lerche and N.~Warner,
  ``Tachyon Condensation on the Elliptic Curve,''
  arXiv:hep-th/0512208.
  %%CITATION = HEP-TH 0512208;%%
}

\lref\HOMFLY{P.~Freyd, D.~Yetter, J.~Hoste, W.~B.~R.~Lickorish, K.~Millett
and O.~ Ocneanu, ``A new polynomial invariant of knots and links,''
Bull.\ Amer.\ Math.\ Soc.\ {\bf 12} (1985) 239;
J.~H.~Przytycki and P.~Traczyk, ``Invariants of links of Conway type,''
Kobe J.\ Math.\ {\bf 4} (1987) 115.}

\lref\kauffman{L.~H.~Kauffman, ``An invariant of regular isotopy,''
Trans.\ Amer.\ Math.\ Soc.\ {\bf 318} (1990) 417.}

\lref\Guadagnini{E~.Guadagnini, ``The Link Invariants of the Chern-Simons
Field Theory: New Developments in Topological Quantum Field Theory,''
Walter de Gruyter Inc., 1997.}

\lref\MOY{H.~Murakami, T.~Ohtsuki, S.~Yamada,
``HOMFLY polynomial via an invariant of colored plane graphs,''
Enseign. Math. {\bf 44} (1998) 325.}

\lref\MuOh{H.~Murakami, T.~Ohtsuki,
``Quantum ${\rm Sp}(n)$ invariant of links via an invariant of colored planar graphs,''
Kobe J. Math. {\bf 13} (1996), no. 2, 191}

\lref\fulton{W.~Fulton and J.~Harris,
``Representation Theory, A First Course,''
Springer, 1991}

\lref\Shumakovitch{A.~Shumakovitch, {\it KhoHo} --- a program for
computing and studying Khovanov homology, {http://www.geometrie.ch/KhoHo}}

\lref\GSV{S.~Gukov, A.~Schwarz and C.~Vafa,
``Khovanov-Rozansky homology and topological strings,''
hep-th/0412243.}

\lref\gornik{B.~Gornik,
``Note on Khovanov link cohomology,''
math.QA/0402266.}

\lref\DGR{ N.~Dunfield, S.~Gukov, J.~Rasmussen, ``The
Superpolynomial for Knot Homologies,'' math.GT/0505662.}

\lref\FloreaM{V.~Bouchard, B.~Florea and M.~Marino, ``Topological
open string amplitudes on orientifolds,'' JHEP {\bf 0502} (2005)
002, hep-th/0411227.}

\lref\ramadevi{
  P.~Borhade and P.~Ramadevi,
  ``SO(N) reformulated link invariants from topological strings,''
  Nucl.\ Phys.\ B {\bf 727} (2005) 471, hep-th/0505008.}

\lref\LPerez{J.~M.~F.~Labastida and E.~Perez,
``A Relation between the Kauffman and the HOMFLY
polynomials for torus knots,'' q-alg/9507031.}

\lref\RKhovanovII{M.~Khovanov, L.~Rozansky, ``Matrix
factorizations and link homology II,'' math.QA/0505056.}

\lref\Kuperberg{G.~Kuperberg, ``The Quantum $G_2$ Link
Invariant,'' math.QA/9201302.}

\lref\Eguchietal{T.~Eguchi, N.~P.~Warner and S.~K.~Yang, ``ADE
singularities and coset models,'' Nucl.\ Phys.\ B {\bf 607} (2001)
3, hep-th/0105194.}

\lref\LercheW{W.~Lerche and N.~P.~Warner, ``Polytopes and solitons
in integrable, N=2 supersymmetric Landau-Ginzburg theories,'
Nucl.\ Phys.\ B {\bf 358} (1991) 571.}

\lref\OShfk{P.~Ozsvath, Z.~Szabo, ``Holomorphic disks and knot invariants,''
Adv. Math. {\bf 186} (2004) 58, math.GT/0209056.}

\lref\Rasmussen{J.~Rasmussen, ``Floer homology and knot complements,''
math.GT/0306378.}

\lref\Rasmussenii{J.~Rasmussen, ``Khovanov homology and the slice genus,''
math.GT/0402131.}

\lref\ESLi{E.~S. Lee, ``The support of the Khovanov's invariants for alternating knots,''
math.GT/0201105.}

\lref\ESLii{E.~S. Lee, ``Khovanov's invariants for alternating links,'' math.GT/0210213.}

%%%%%
\def\boxit#1{\vbox{\hrule\hbox{\vrule\kern8pt
\vbox{\hbox{\kern8pt}\hbox{\vbox{#1}}\hbox{\kern8pt}}
\kern8pt\vrule}\hrule}}
\def\mathboxit#1{\vbox{\hrule\hbox{\vrule\kern8pt\vbox{\kern8pt
\hbox{$\displaystyle #1$}\kern8pt}\kern8pt\vrule}\hrule}}

%%%%%%%%%%%%%%%%%%%%%%%%%%%  FIGURES   %%%%%%%%%%%%%%%%%%%%%%%%%%%%%%%

\let\includefigures=\iftrue
\newfam\black
\includefigures
\input epsf
\def\figin{\epsfcheck\figin}\def\figins{\epsfcheck\figins}
\def\epsfcheck{\ifx\epsfbox\UnDeFiNeD
\message{(NO epsf.tex, FIGURES WILL BE IGNORED)}
\gdef\figin##1{\vskip2in}\gdef\figins##1{\hskip.5in}% blank space instead
\else\message{(FIGURES WILL BE INCLUDED)}%
\gdef\figin##1{##1}\gdef\figins##1{##1}\fi}
\def\DefWarn#1{}
\def\figinsert{\goodbreak\midinsert}
\def\ifig#1#2#3{\DefWarn#1\xdef#1{fig.~\the\figno}
\writedef{#1\leftbracket fig.\noexpand~\the\figno}%
\figinsert\figin{\centerline{#3}}\medskip\centerline{\vbox{\baselineskip12pt
\advance\hsize by -1truein\noindent\footnotefont{\bf Fig.~\the\figno:} #2}}
\bigskip\endinsert\global\advance\figno by1}
%%%
\else
\def\ifig#1#2#3{\xdef#1{fig.~\the\figno}
\writedef{#1\leftbracket fig.\noexpand~\the\figno}%
%\figinsert\figin{\centerline{#3}}\medskip\centerline{\vbox{\baselineskip12pt
%\advance\hsize by -1truein\noindent\footnotefont{\bf Fig.~\the\figno:} #2}}
%\bigskip\endinsert
\global\advance\figno by1}
\fi

%%%%%%%%%%%%%%%%%%%%%%%%%%%  YOUNG TABLEUAS  %%%%%%%%%%%%%%%%%%%%%%%%%%%%%%%
%%
%%                              TABLEAUX.TEX
%%      This  macro file is for producing a ``Young Tableau'' which is
%%      an array of little squares sometimes used in mathematical physics.
%%      For instance, the command $\tableau{6 3 2}$ will produce a tableau
%%      with 6 squares in the top row, 3 in the next, and 2 in the last.
%%                                  OOOOOO
%%      This tableau will look like OOO    but made of squares instead of O's.
%%                                  OO
%%      Any number of rows may be present, each having a nonzero number of
%%      squares.
%%
%%      A tableau is math mode material, so use $ or $$ to enclose it.
%%
%%      The size and line-thickness of the little boxes are controlled by the
%%      dimension parameters --
%%              \tableauside=1.0ex              %(size)
%%              \tableaurule=0.4pt              %(line-thickness)
%%      Change them if you want.
%%
%%                                                      -- Doug Eardley 9/19/8%%
%%
\newdimen\tableauside\tableauside=1.0ex
\newdimen\tableaurule\tableaurule=0.4pt
\newdimen\tableaustep
\def\phantomhrule#1{\hbox{\vbox to0pt{\hrule height\tableaurule width#1\vss}}}
\def\phantomvrule#1{\vbox{\hbox to0pt{\vrule width\tableaurule height#1\hss}}}
\def\sqr{\vbox{%
  \phantomhrule\tableaustep
  \hbox{\phantomvrule\tableaustep\kern\tableaustep\phantomvrule\tableaustep}%
  \hbox{\vbox{\phantomhrule\tableauside}\kern-\tableaurule}}}
\def\squares#1{\hbox{\count0=#1\noindent\loop\sqr
  \advance\count0 by-1 \ifnum\count0>0\repeat}}
\def\tableau#1{\vcenter{\offinterlineskip
  \tableaustep=\tableauside\advance\tableaustep by-\tableaurule
  \kern\normallineskip\hbox
    {\kern\normallineskip\vbox
      {\gettableau#1 0 }%
     \kern\normallineskip\kern\tableaurule}%
  \kern\normallineskip\kern\tableaurule}}
\def\gettableau#1 {\ifnum#1=0\let\next=\null\else
  \squares{#1}\let\next=\gettableau\fi\next}

\tableauside=1.0ex
\tableaurule=0.4pt

%%%%%%%%%%%%%%%%%%%%%  Math-style letters   %%%%%%%%%%%%%%%%%%%%%%%%
\font\cmss=cmss10 \font\cmsss=cmss10 at 7pt

\def\IB{\relax\hbox{$\inbar\kern-.3em{\rm B}$}}
\def\IC{\relax\hbox{$\inbar\kern-.3em{\rm C}$}}
\def\IQ{\relax\hbox{$\inbar\kern-.3em{\rm Q}$}}
\def\ID{\relax\hbox{$\inbar\kern-.3em{\rm D}$}}
\def\IE{\relax\hbox{$\inbar\kern-.3em{\rm E}$}}
\def\IF{\relax\hbox{$\inbar\kern-.3em{\rm F}$}}
\def\IG{\relax\hbox{$\inbar\kern-.3em{\rm G}$}}
\def\IGa{\relax\hbox{${\rm I}\kern-.18em\Gamma$}}
\def\IH{\relax{\rm I\kern-.18em H}}
\def\IK{\relax{\rm I\kern-.18em K}}
\def\IL{\relax{\rm I\kern-.18em L}}
\def\IP{\relax{\rm I\kern-.18em P}}
\def\IR{\relax{\rm I\kern-.18em R}}
\def\Z{\relax\ifmmode\mathchoice
{\hbox{\cmss Z\kern-.4em Z}}{\hbox{\cmss Z\kern-.4em Z}}
{\lower.9pt\hbox{\cmsss Z\kern-.4em Z}}
{\lower1.2pt\hbox{\cmsss Z\kern-.4em Z}}\else{\cmss Z\kern-.4em
Z}\fi}
\def\IZ{Z\!\!\!Z}
\def\II{\relax{\rm I\kern-.18em I}}

\def\S{{\bf S}}

\def\R{{\bf R}}

%%%%%%%%%%%%%%%%%%%%% Calligraphic letters  %%%%%%%%%%%%%%%%%%%%%

\def\CC {{\cal C}}

\def\CF {{\cal F}}

\def\CH {{\cal H}}

\def\CJ {{\cal J}}

\def\CL {{\cal L}}

\def\CO {{\cal O}}
\def\CP {{\cal P}}

%%%%%%%%%%%%%%%%%%%%%%%%%% Derivatives  %%%%%%%%%%%%%%%%%%%%%%%%

\def\p{\partial}

%%%%%%%%%%%%%%%%%%%% letters with bar %%%%%%%%%%%%%%%%%%%%%%%%%%
\def\tilde{\widetilde}
\def\hat{\widehat}
\def\bar{\overline}

%%%%%%%%%%%%%%%%%%%%%%%%%%% Math symbols %%%%%%%%%%%%%%%%%%%%%%%

\def\p{\partial}

\def\lieg{{\bf g}}

\def\inbar{\,\vrule height1.5ex width.4pt depth0pt}

%%%%%%%%%%%%%%%%%%%   Greek letters %%%%%%%%%%%%%%%%%%%

\def\la{\lambda}

\def\bar{\overline}

\def\det{{\rm det}}

\def\IH{{\bf H}}

\def\example#1{\bgroup\narrower\footnotefont\baselineskip\footskip\bigbreak
\hrule\medskip\nobreak\noindent {\bf Example}. {\it #1\/}\par\nobreak}
\def\endexample{\medskip\nobreak\hrule\bigbreak\egroup}

%%%
%%%%%% Conventions
%%%%%%%%%

   % Poincare polynomial of reduced SL(N) homology
 % unreduced SL(N) homology
\def\SH{{\CP}}         % reduced HOMFLY homology
\def\bSH{\bar{\CP}}    % unreduced HOMFLY homology

\def\HSO{{\it HSO}}       % Poincare polynomial of reduced SO(N) homology
\def\bHSO{\bar {\it HSO}}      % unreduced SO(N) homology
\def\HSp{{\it HSp}}
\def\bHSp{\bar {\it\HSp}}

\def\SK{{\CF}}         % reduced Kauffman
\def\bSK{\bar{\CF}}    % unreduced Kauffman

\def\du{d_{\rightarrow}}  % universal differentials
\def\duu{d_{\leftarrow}}
\def\duuu{d_{\rlap{\raise 1.2pt\hbox{$\scriptstyle\rightarrow$}}\lower 1.2pt
\hbox{$\scriptstyle\leftarrow$}}}
\def\Qu{Q_{\rightarrow}}
\def\Quu{Q_{\leftarrow}}
\def\leadsto{\rightsquigarrow}

%%%%%%%%%%%%%%%%%%% TITLE PAGE  %%%%%%%%%%%%%%%%%%%%%%%%%%%%%%%%

\Title{\vbox{\baselineskip11pt
\hbox{hep-th/0512298}
\hbox{CALT-68-2584}
\hbox{NSF-KITP-05-123}
}}
{\vbox{
\centerline{Matrix Factorizations and Kauffman Homology}
}}
\centerline{Sergei Gukov$^{a,c}$ and Johannes Walcher$^{b,c}$}
\medskip
\medskip
\medskip
\vskip 8pt
\centerline{$^a$ \it California Institute of Technology 452-48,
Pasadena, CA 91125, USA}
\medskip
\centerline{$^b$ \it Institute for Advanced Study, Princeton, NJ 08540, USA}
\medskip
\centerline{$^c$ \it Kavli Institute for Theoretical Physics, Santa Barbara, 
CA 93106, USA}
\medskip
\medskip
\medskip
\noindent

\vskip 20pt {\bf \centerline{Abstract}} \noindent

The topological string interpretation of homological knot invariants has led 
to several insights into the structure of the theory in the case of $sl(N)$.
We study possible extensions of the matrix factorization approach to knot 
homology for other Lie groups and representations. In particular, we introduce 
a new triply graded theory categorifying the Kauffman polynomial, test it, 
and predict the Kauffman homology for several simple knots.

\smallskip

\medskip
\Date{December 2005}

%\listtoc\writetoc

%%%%%%%%%%%%%%%%%%%%%%%%%%%%%%%%%%%%%%%%%%%%%%%%%%%%%%%%%%%%%

\newsec{Introduction}

In this paper, we study from the physical perspective knot invariants of the 
homological type. The exciting progress in this relatively young mathematical 
field opens many new directions for  the  interaction between different branches 
of mathematics and physics, such as gauge theory, topological string theory, 
symplectic geometry, representation theory, and low-dimensional topology.
The present status of the field suggests several lines of  connection, which include
the appearance of matrix factorizations in the definition of Khovanov-Rozansky theory
and the evidence for a triply-graded theory unifying various knot  homologies.
The purpose of the present work is to develop both of these connections and to present
further evidence for the existence of a much richer underlying structure.

Recall that Chern-Simons gauge theory, in which polynomial knot invariants are obtained as 
expectation values of Wilson lines \WittenJones, is connected to topological strings
on Calabi-Yau manifolds.
Building on earlier work of Witten \Wittencsstring,
Gopakumar and Vafa \refs{\GopakumarV,\GViii} proposed a relation between Chern-Simons theory
on the three-sphere and the closed topological string on a particular Calabi-Yau three-fold.
In this setup, knots can be incorporated by introducing D-branes in the closed string geometry,
so that the corresponding polynomial invariants are related to topological string amplitudes
in the D-brane background \OV.
Alternatively, they can be viewed as generating functions that
count dimensions of Hilbert spaces of BPS states in the physical
string theory with plus-minus signs \refs{\OV,\LMV}.

As pointed out in \GSV, this interpretation is conceptually similar to a ``categorification'',
which is a lift of a polynomial invariant to a homology theory whose graded Euler 
characteristic is the polynomial at hand. Over the past years, several such homological 
knot invariants have been discovered, including Khovanov homology \Khovanov\ whose Euler 
characteristic is the Jones polynomial, knot Floer homology \refs{\OShfk,\Rasmussen} whose 
associated classical invariant is the Alexander polynomial, Khovanov-Rozansky homology 
\refs{\Khovanoviii,\RKhovanov} which categorifies the quantum $sl(N)$ invariant, and 
several others. The connection between these homological invariants and the topological 
string has led to many new predictions and insights into the structure of the theory.
For example, it was found \refs{\GSV,\DGR}, that the invariants  associated with knots
decorated with the fundamental representation of $sl(N)$ should unify at large $N$ into 
a single, triply-graded structure, with well-controlled deviations at small $N$.
This triply graded theory is a categorification of one of the two-variable polynomial
invariants, the HOMFLY polynomial \HOMFLY. (Another, conceivably related, categorification 
of the HOMFLY polynomial was proposed in \RKhovanovII.)

The physical picture suggests the existence of homological knot invariants
associated with a large class of Lie algebras and representations.
It is likely that many of them can be constructed using matrix factorizations, as
in \refs{\RKhovanov,\RKhovanovII}.
Using the connection to the physics of two-dimensional 
Landau-Ginzburg models, we will argue in this paper that this is indeed the case.
In particular, we will present a list of Landau-Ginzburg potentials
which can be used for the construction of homological invariants associated
with a large class of Lie  algebras and representations.
Some of these potentials are new. The connection to Landau-Ginzburg models 
will also be useful to us for understanding relations between different knot 
homologies.

In the topological string picture, we can introduce
orientifolds to naturally define invariants associated with the 
fundamental representation of the other classical Lie algebras with a 
large $N$ expansion, $so(N)$ and $sp(N)$. This modification was used
to study the Chern-Simons partition function in \SinhaV,
and the polynomial knot invariants in \refs{\FloreaM,\ramadevi}.
At this level, the $so(N)/sp(N)$ invariants are known to unify into the Kauffman 
polynomial \kauffman,
which is the second two-variable polynomial knot invariant. We will argue that
this can be lifted to the homological level as well. 
Moreover, we will present evidence based on the Landau-Ginzburg picture
that the resulting triply-graded homology theory can not only be reduced to the homological 
invariants associated with the fundamentals of $so(N)/sp(N)$,
but in fact also contains the triply graded HOMFLY homology.
In this sense, the Kauffman homology we are proposing in this paper
might be the most fundamental homological invariant to date.

The paper is organized as follows. We start in section 2 with a summary
of notations and definitions. In section 3, we explain the role of Landau-Ginzburg
potentials in knot homology and use the intuition from physics to write
new potentials associated with various representations of classical
Lie algebras. We further use the relation with Landau-Ginzburg theories
in section 4 to study various properties of knot homologies.
In particular, we find families of differentials associated with deformations
of Landau-Ginzburg potentials.
In section 5, we discuss physical interpretation of $so(N)/sp(N)$ knot homologies
in the context of topological strings with an orientifold.
As in the $sl(N)$ case, this interpretation leads us to a new triply-graded
theory that unifies $so(N)/sp(N)$ knot homologies for all $N$. 
The study of this theory is the subject of section 6.

%%%%%%%%%%%%%%%%%%%%%%%%%%%%%%%%%%%%%%%%%%%%%%%%%%%%%%%%%%%%

\newsec{Preliminaries}

\noindent Our notations are summarized in the following table:

\halign{\quad # & \vtop{\parindent=0pt\hsize=25em\hangindent.5em\strut#\strut} \cr
 $K$ & a knot (a link) \cr
 $\lieg$ & semisimple Lie algebra\cr
 $R$ & a representation of $\lieg$ \cr }

\halign{\quad # & \vtop{\parindent=0pt\hsize=32em\hangindent.5em\strut#\strut} \cr
\noalign{\noindent {\bf Polynomial invariants}}
\noalign{\noindent two-variables polynomials}
$P (\la,q)$ & normalized HOMFLY polynomial \cr
$F (\la,q)$ & normalized Kauffman polynomial \cr
\noalign{\noindent one-variable polynomials (``quantum'' invariants)}
$\bar P^{\lieg,R}(q)$ & quantum invariant associated with Lie algebra $\lieg$ and
representation $R$ \cr
$P_N(q)$, $\bar P_N (q)$ & normalized/unnormalized quantum $sl(N)$ invariant ($\lieg=sl(N)$,
$R=\tableau{1}$) \cr
$J_n(q)$ & colored Jones polynomial ($\lieg=sl(2)$, $R=$$n$-dimensional representation) \cr
$F_N(q)$, $\bar F_N (q)$ & quantum $so(N)$ invariant \cr
$C_N(q)$, $\bar C_N (q)$ & quantum $sp(N)$ invariant \cr}

\vskip 5pt

\halign{\quad #& \quad # & \vtop{\parindent=0pt\hsize=23em\strut#\strut} \cr
\noalign{\noindent{\bf Homological invariants and their Poincar\'e polynomials}}
\noalign{\noindent triply-graded}
$\CH^{\rm HOMFLY}$ & $\SH(\la,q,t)$, $\bSH(\la,q,t)$
& triply-graded theory and reduced/unreduced superpolynomial from \DGR \cr
$\CH^{\rm Kauffman}$ & $\SK(\la,q,t)$, $\bSK(\la,q,t)$
& triply-graded Kauffman theory and its reduced/un\-re\-duced superpolynomial \cr
\noalign{\noindent doubly-graded}
$\CH^{\lieg,R}$ & $\CP^{\lieg,R}$ & homological invariant associated with $\lieg$ and $R$ \cr
$\CH^{sl(N),\tableau{1}}$ & $KhR_N(q,t)$  & Khovanov-Rozansky $sl(N)$ homology \cr
$\CH^{sl(2),\tableau{1}}$ & $Kh(q,t)$ & Khovanov $sl(2)$ homology \cr
$\CH^{so(N),\tableau{1}}$ & $\HSO_N(q,t)$ & $so(N)$ homology \cr
$\CH^{sp(N),\tableau{1}}$ & $\HSp_N(q,t)$ & $sp(N)$ homology \cr}

\vskip 5pt
%\noindent{{\bf Table 1:} Summary of notation for knot invariants.
Here, we have assumed the underlying knot to be fixed, which if need arises, we include
as an additional variable, as in $P(K;\la,q)$. As a general rule, this
notation refers to the {\it normalized} version of the invariants (wherever this
notion applies). When appropriate, the {\it unnormalized} version will denoted by
over-lining it.

Let us explain our conventions in more detail.

\noindent{\bf Two-variable polynomials:} The {\it unnormalized} HOMFLY polynomial,
$\bar P(K;\la,q)$ is the polynomial invariant of unoriented knots in $S^3$ defined
by the skein relations of oriented planar diagrams
\eqn\skeinp{ \la \bar P (\lower3.0pt
\hbox{\epsfxsize0.15in\epsfbox{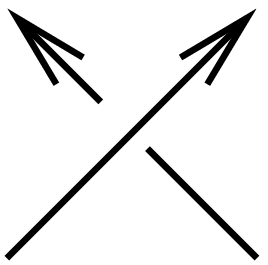}})
- \la^{-1} \bar P(\lower3.0pt
\hbox{\epsfxsize0.15in\epsfbox{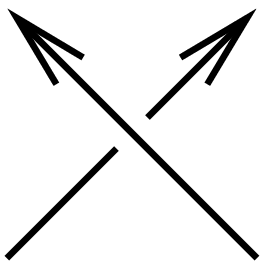}})
=
(q-q^{-1}) \bar P(\lower3.0pt
\hbox{\epsfxsize0.15in\epsfbox{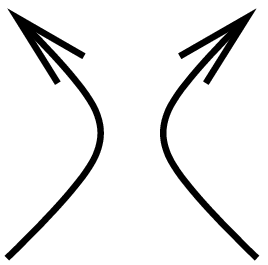}})
}
with the normalization
\eqn\punknot{\bar P ({\rm unknot}) = {\la - \la^{-1} \over q -
q^{-1} } }
The {\it unnormalized} Kauffman polynomial $\bar F (K;\la,q)$ is another
invariant of unoriented knots which is defined by a similar set of combinatorial
rules. We first define an invariant of planar diagrams $\tilde{F}(L;\la,q)$
via the skein relations
\eqn\skeinf{\eqalign{ & \tilde F ({\lower1.0pt
\hbox{\epsfxsize0.3in\epsfbox{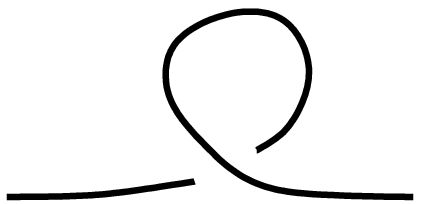}}}) = \la \tilde F (-)
\qquad\qquad
\tilde F({\lower1.0pt
\hbox{\epsfxsize0.3in\epsfbox{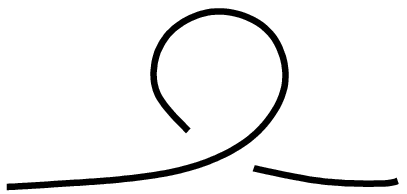}}}) = \la^{-1} \tilde F (-)
\cr
 & \tilde F ({\lower3.0pt
\hbox{\epsfxsize0.15in\epsfbox{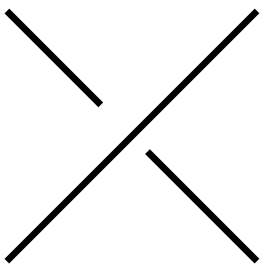}}}) - \tilde F
({\lower3.0pt \hbox{\epsfxsize0.15in\epsfbox{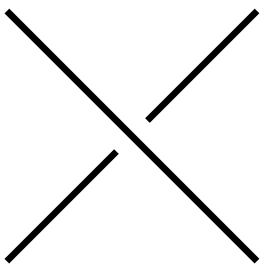}}}) =
(q-q^{-1}) (\tilde F ({\lower3.0pt
\hbox{\epsfxsize0.15in\epsfbox{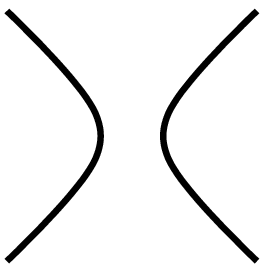}}}) - \tilde F
({\lower3.0pt \hbox{\epsfxsize0.15in\epsfbox{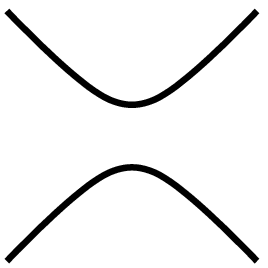}}})) }}
and the normalization
\eqn\funknot{\tilde F ({\rm unknot}) = {\la - \la^{-1} \over q -
q^{-1} } + 1 }
Then, if $w(K)=$(number of ``$+$'' crossings)$-$(number
of ``$-$'' crossings) is the writhe of $K$,
the Kauffman polynomial is given by
\eqn\kauff{ \bar F(K;\la,q) = \la^{-w(K)} \tilde F(K;\la,q)}
The normalized versions of the HOMFLY and Kauffman polynomials,
$P(K)$ and $F(K)$, can be defined by the same combinatorial rules,
with the normalization where the unknot evaluates to 1. In
particular, we have
\eqn\pnorm{\eqalign{\bar P (K) &= \bar P ({\rm unknot}) P(K) \cr
\bar F (K) &= \bar F ({\rm unknot}) F(K) }}

\medskip\noindent
{\bf Remark:} In the knot theory literature (and, {\it e.g.}, the {\it Mathematica}
Package {\it KnotTheory}), the HOMFLY and Kauffman polynomials are often expressed
in terms of variables $a$ and $z$, in conventions which are related to ours via
\eqn\knottheory{\eqalign{
P_{\rm ours}(\la,q) &= P_{\rm KnotTheory}(a=\la,z=q-q^{-1}) \cr
F_{\rm ours}(\la,q) &= F_{\rm KnotTheory}(a=i\la,z=-i(q-q^{-1}))
}}

\medskip\noindent
{\bf Symmetries:}
The HOMFLY and Kauffman polynomial have certain symmetries.
Let us denote by $\bar K$ the mirror image of the knot $K$, which is obtained
by exchanging positive with negative crossings. We have
\eqn\symmetries{\eqalign{
P(\bar K;\la^{-1},q^{-1}) &= P(K;\la,q) = P(K;-\la,q^{-1})=P(K;-\la,q) \cr
F(\bar K;\la^{-1},q^{-1}) &= F(K;\la,q) = F(K;-\la,q^{-1})}}

\noindent{\bf Quantum Invariants:}
The (unnormalized) quantum invariants $P^{\lieg,R}$ are best defined as
expectation values of Wilson loop operators in representation $R$ in
Chern-Simons theory based on Lie algebra $\lieg$,
\eqn\wilson{\bar P^{\lieg,R}(q) = \langle W_{R} (K) \rangle_{\lieg,k}}
where, the level $k$ of the Chern-Simons theory is related to $q$ via
\eqn\level{q = e^{\pi i\over k+ h}}
where $h$ is the dual Coxeter number of $\lieg$.

\noindent{\bf Specializations:}
As is well-known \Turaev, the quantum invariants for the fundamental representations
of the classical Lie algebras can be obtained as specializations of the 2-variable
polynomials, $F$ and $P$. With the above conventions, we have
\eqn\specializations{
\eqalign{ sl(N):& \qquad P_N(q)  =P(\la=q^N,q) = P_{\rm KnotTheory}(a=q^{N},z=q-q^{-1}) \cr
so(N):& \qquad F_N(q) = F(\la=q^{N-1},q) = F_{\rm KnotTheory}(i q^{N-1}, -i (q-q^{-1})) \cr
sp(N):& \qquad C_N(q) = F(\la=-q^{N+1},q) = F_{\rm KnotTheory}(i q^{N+1}, i (q-q^{-1})) \cr
}}
where in the $sp$ case $N$ is even and the rank of it is $N/2$.
In \Turaev, another specialization is mentioned which is related
to the quantum invariant associated with twisted Kac-Moody algebra
$A_{N-1}^{(2)}$; it is given by
\eqn\another{
F(q^{N-1},q) = F_{\rm KnotTheory}(iq^{N-1},-i(q-q^{-1}))
}

\noindent
{\it Special cases:} The relation to the classical Jones polynomial
(in the {\it KnotTheory} conventions) is
$$
P_2(q) = J(q^{-2})
$$
The isomorphism $so(4)\cong sl(2)\times sl(2)$ yields the relation
\eqn\sofourpolrel{ F_4(q) = P_2(q)^2 }
The isomorphism $sp(2) \cong sl(2)$ yields
\eqn\sptwopolrel{ C_2(q) = P_2(q^2) }
which is equivalent to the classical relation between the
Jones and the Kauffman polynomial,
$$
J(q^{-4}) = J((iq)^{-4}) = P_2(q^2) = C_2(q)
= F(iq^3,i(q-q^{-1})) = F(-q^3,q+q^{-1})
$$
Finally, we comment that neither $so(6)\cong su(4)$ nor
$sp(4)\cong so(5)$ yield any particular kind of relationship
because different representations are involved.

Let us also note that in virtue of the symmetry relations \symmetries,
the specializations \specializations\ imply the relationship
\eqn\negativeN{\eqalign{
C_N(\bar K;q) &= F(K;\la=q^{-N-1},q) \cr
F_N(\bar K;q) &= F(K;\la=-q^{-N+1},q) }}
In other words, the continuation of $so(N)/sp(N)$ invariants to negative
$N$ can be viewed as the $sp(N)/so(N)$ invariant {\it for the mirror
knot}. We will interpret this in the physical setup in section 5.

%%%%%%%%%%%%%%%%%%%%%%%%%%%%%%%%%%%%%%%%%%%%%%%%%%%%%%%%%%%%

\newsec{ABDE of Matrix Factorizations}

As we already mentioned in the introduction, we believe that the
categorification of polynomial knot invariants can be extended
to a much larger class of invariants associated with different Lie
algebras and representations than what has been considered so far.
The motivation for this comes, on one hand, from the realization of
knot homologies in topological string theory which will be the subject
of section 5, and, on the other hand, from the connection with
Landau-Ginzburg models which will be discussed here.
For some further aspects of the relation between matrix factorizations
and the physics of Landau-Ginzburg models, see refs.\ 
\refs{\Brunner,\KapustinLii,\Emanuelii,\calin,\HoriJW,\Brunneriii,\Rstability,
\GovindarajanIM}.

We denote the homology theory associated with
a simple Lie algebra $\lieg$ and a representation $R$ 
by $\CH_{i,j}^{\lieg,R} (K)$, or simply by $\CH_{i,j}^{\lieg} (K)$ when
a particular representation is clear from the context.
The Poincare polynomial of this theory is denoted $\bar \CP^{\lieg,R}$,
and the graded Euler characteristic is equal to the quantum group
invariant
\eqn\hgreuler{
\bar P^{\lieg,R} (q) = \sum_{i,j \in \Z} (-1)^i q^j \dim \CH_{i,j}^{\lieg,R} (K) }
When $R$ is a $N$-dimensional vector representation of $sl(N)$ (resp.\
$so(N)$ or $sp(N)$) we refer to $\CH_{i,j}^{\lieg,R} (K)$ as the $sl(N)$ 
(resp.\ $so(N)$ or $sp(N)$) knot homology. We often denote the
vector representation as $V$ and the spinor representation (of
$so(N)$) as $S$.
We hope that some of the knot homologies $\CH_{i,j}^{\lieg,R} (K)$
can be constructed using matrix factorizations, as in \refs{\RKhovanov,\RKhovanovII}.

Let us briefly recall the construction of the Khovanov-Rozansky homology
for the fundamental representation of $sl(N)$ \RKhovanov.
Given a planar diagram of a knot (or link, or tangle), in this construction
one defines a certain complex of matrix factorizations of (mostly degenerate)
potentials in variables associated with the components of the planar diagram.
The cohomology of this $\Z\oplus\Z\oplus\Z_2$ graded complex
defines $\CH_{i,j}^{sl(N)} \equiv \CH_{i,j}^{sl(N), \tableau{1}}$.
For example, to a single crossing-less line starting at a point labeled
$x$ and ending at a point labeled by $y$ one associates the potential
$x^{N+1}-y^{N+1}$ and the factorization $(x-y)\pi_{x,y}=x^{N+1}-y^{N+1}$
for the appropriate choice of $\pi_{x,y}$. To compute the homology of the
simplest knot, the unknot, we identify the two ends of this line.
This leads us to the factorization $0\cdot x^N=0$ of the trivial potential.
The cohomology of the two-periodic complex
\def\mapright#1{\mathop{\longrightarrow}\limits^{#1}}
\eqn\unknot{
C({\rm unknot})=\bigl(\cdots\mapright{ }\IC[x]\mapright{0} \IC[x]
\mapright{x^N} \IC[x]\mapright{ } \cdots\bigr)}
is just the Jacobi ring of the potential $W_{sl(N), \tableau{1}} = x^{N+1}$,
\eqn\hslnunknot{\eqalign{
\CH_{i,j}^{sl(N)}({\rm unknot}) = & H(C({\rm unknot})) \cr
= & \CJ(x^{N+1}) \cr
= & \IC[x]/ x^N \cr
= & \{1,x,\ldots,x^{N-1}\}
}}
The polynomial grading of these homology groups is shifted down
by $N-1$ units, so that the Poincare polynomial of the unknot is
$$
\bar{KhR}_N({\rm unknot}) = q^{-N+1} + q^{-N+3} +\cdots+q^{N+1}
= {q^N-q^{-N} \over q-q^{-1}}
$$
The knot homology of the unknot in this case can also be interpreted
as the cohomology ring of complex projective space ${\bf CP}^{N-1}$.
It has been suggested in \RKhovanov\ that the extension of this result to
the $k$-th antisymmetric representation $\Lambda^k V$ is given by
the cohomology ring of the Grassmannian\foot{Sigma-model with target
space ${\it Gr}(k,N)$ appears as a theory on the intersection of
$k$ compact D-branes with $N$ non-compact D-branes in the string
theory realization.} of $k$-planes in $\IC^N$:
\eqn\grass{
\CH_*^{sl(N), {{\tableau{1 1}\atop{\cdot}}\atop\tableau{1}}\raise
2pt\hbox{$\Bigr\} \scriptstyle k$}}
({\rm unknot}) = H^* ({\it Gr}(k,N))}
and that the generalization of the matrix factorization construction
to these representations should be based on the multi-variable potential
\eqn\Wanti{W_{sl(N),\Lambda^k}(z_1,\ldots,z_k) = x^{N+1}_1+\cdots+x_k^{N+1}}
The right-hand side of this expression should be viewed as a function
of the variables $z_i$, which are the elementary symmetric polynomials in the $x_j$,
$$
z_i = \sum_{j_1<j_2<\cdots<j_i} x_{j_1}x_{j_2}\cdots x_{j_i}
$$
In this case, the matrix factorization associated with the
unknot is the tensor product of factorizations
$0\cdot\partial_{z_i} W_{sl(N),\Lambda^k}$ over $i=1,2,\ldots,k$.
Its cohomology is again the Jacobi ring of the potential
$W_{sl(N),\Lambda^k}$, which is well-known to be the
cohomology of the Grassmannian $H^* ({\it Gr}(k,N))$.

To summarize, one of the key elements in this approach is
the problem of constructing a potentials $W_{\lieg,R} (x_i)$
for given $\lieg$ and $R$, such that the Jacobi ring of $W_{\lieg,R}$
is isomorphic to the homology of the unknot
\eqn\jhgeneralgr{ \CH^{\lieg,R}({\rm unknot}) \cong \CJ (W_{\lieg,R} (x_i)) }
This isomorphism involves a shift of grading (by $N-1$ units in the case
of $sl(N)$), which corresponds to the spectral flow from NS to Ramond sector.
At present, $W_{\lieg,R} (x_i)$ is known only in the special cases we
mentioned above: the fundamental and the totally antisymmetric
representations of $sl(N)$.
In what follows, our goal will be to expand this list and to derive
potentials for other Lie algebras and representations using insights
from conformal field theory and topological strings.
In particular, we shall think of $W_{\lieg,R} (x_i)$ as the potential
in the topological Landau-Ginzburg model, in which topological
D-branes are described by matrix factorizations \refs{\Brunner,\KapustinLii}.
Of course, not all ring relations can be derived from a potential,
and we will give an example of this below.

Another key elements in the construction of \RKhovanov\
is a set of combinatorial rules (skein relations) which allow
us to define homological invariants of planar graphs and,
eventually, knots and links. Unfortunately, even for polynomial knot 
invariants such skein relations are not known except in special cases. 
If $R$ is a spinor or a vector representation
of $so(N)$ or $sp(N)$, it is natural to start with the skein relations for the
Kauffman polynomial discussed above. Sometimes these skein relations
can be simplified further to a set of rules analogous to the
Murakami-Ohtsuki-Yamada rules \MOY\ for $\lieg = sl(N)$. For example, for
$so(5)$ such rules were constructed by Kuperberg \Kuperberg.
Similarly, for $so(6) \cong sl(4)$ the spinor (resp. vector) of $so(6)$ is
identified with the vector (resp. antisymmetric) representation of $sl(4)$.
Hence, the $so(6)$ calculus is identical to the usual MOY calculus for 
$su(4)$. A set of rules also exists for the fundamental of $sp(N)$ \MuOh.

\ifig\knothoma{Planar trivalent graph near a wide edge.}
{\epsfxsize3.0in\epsfbox{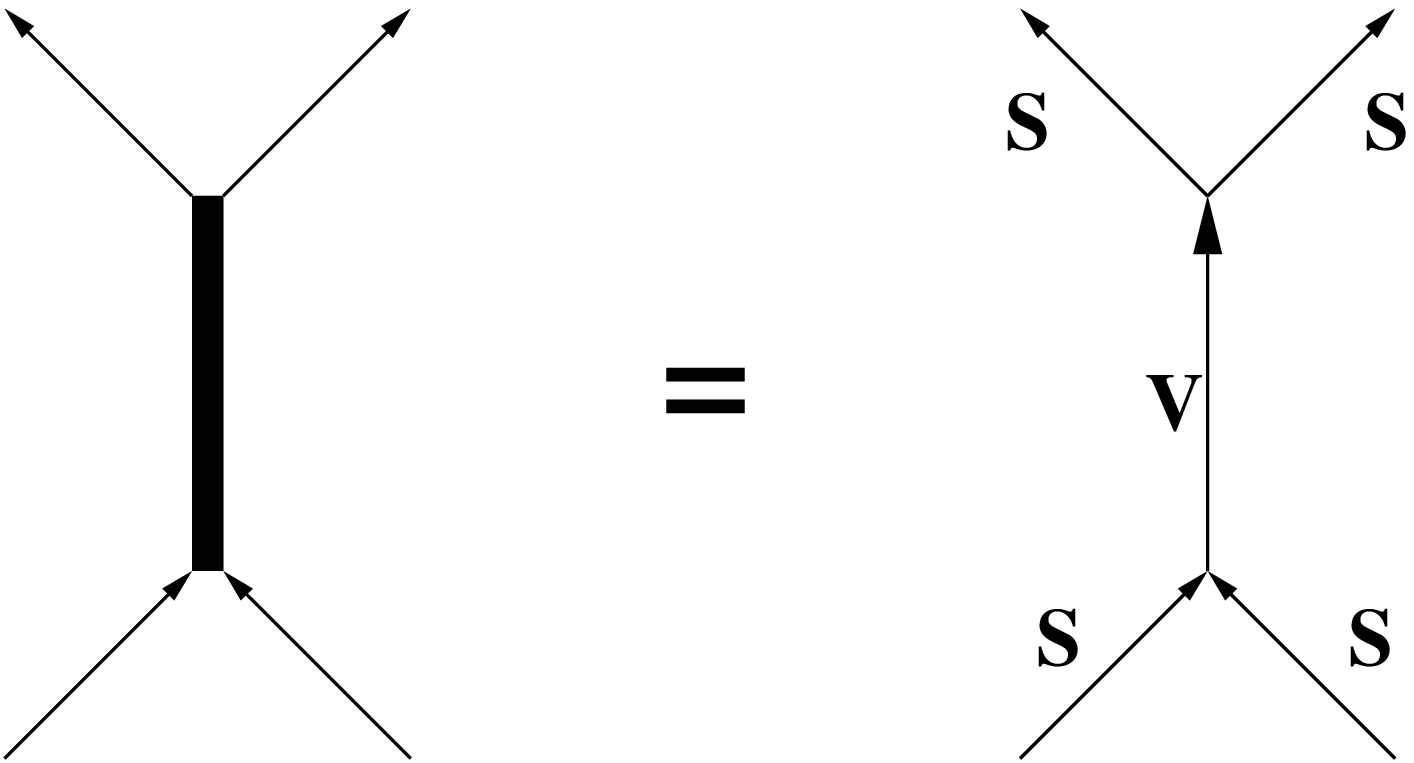}}

Motivated by these special cases, it is natural to expect that in the $so(N)$ 
case for general $N$, a knot diagram can be reduced to (sums of) planar graphs
built out of two types of edges, corresponding to the spinor and vector
representations, respectively, see \knothoma:
\eqn\soedges{\eqalign{ & S~:~~ {\rm ~thin~edge~} \cr & V~:~~ {\rm
~wide~edge~} }}
By analogy with the $sl(N)$ case studied in \refs{\RKhovanov},
to thin/wide edges we associate, respectively, potentials $W_{so(N),S}$
and $W_{so(N),V}$ (see below). More precisely, we have sums of those
potentials for all incoming and outgoing edges. It should then
be possible to find an appropriate matrix factorization corresponding
to such a planar diagram, as well as to concatenate those factorizations
into the complex which will compute $\CH^{so(N)}$ for an arbitrary
knot or link.\foot{For $N=5$, one can try to construct the $so(5)$ knot
homology using web cobordisms, following the steps of \Khovanoviii.
In this construction, web cobordisms should have different types of
edges, faces, and dots.}

\example{$so(6)$} The potential for the spinor of $so(6)$
(fundamental of $sl(4)$) is $W_{so(6),S} = x^5$.
Let us consider a tensor product of two such representations.
It corresponds to the potential
$$
\eqalign{
W_{so(6),S} (x_1) + W_{so(6),S} (x_2)
%x_1^5 + x_2^5
& = (x_1+x_2)^5 - 5 x_1x_2 (x_1+x_2)^3 + 5
(x_1x_2)^2 (x_1+x_2) \cr & = z^5 - 5 y z^3 + 5 y^2 z }
$$
where $z = x_1+x_2$ and $y = x_1 x_2$. After a linear change of
variables, this is the same as $W_{so(6),V} = z^5 + z y^2$.
\endexample

%%%%%%%%%%%%%%%%%%%%%%%%%%%%%%%%%%%

\subsec{Knot Homologies and Hermitian Symmetric Spaces}

One of our main examples is the $so(N)$ knot homology, that is the
case of $\lieg = so(N)$ and $R = V$. This theory should be a
categorification of the quantum $so(N)$ invariant introduced in the
previous section,
\eqn\fnviah{ \bar F_N (q) = \sum_{i,j \in \Z} (-1)^i q^j \dim
\CH_{i,j}^{so(N)} (K) }
Using the physical picture, we shall see that the corresponding potential is
\eqn\wfordn{ W_{so(N)} = x^{N-1} + xy^2 }
Indeed, this potential leads to the correct homology of the unknot,
which is isomorphic to the Jacobi ring of the D$_N$ singularity,
\eqn\dnunknot{ \CH_*^{so(N)} ({\rm unknot}) = \IQ[x,y] / \{
x^{N-2}+y^2,xy \} }
It is $N$-dimensional, in agreement with \funknot, generated by
$x$ of $(q,t)$-degree $(2,0)$ and $y$ of degree $(N-2,0)$. The
homology \dnunknot\ is also isomorphic to the cohomology of
the homogeneous coset space
\eqn\socoset{{SO(N)\over SO(N-2)\times U(1)}}
This relation and the relation between the unknot homology for
the anti-symmetric representations of $sl(N)$ and the cohomology of the
Grassmannian \grass\ is suggestive of the following generalization.
There is a well-known relationship between the chiral rings of certain
Landau-Ginzburg potentials, the cohomology of compact hermitian
symmetric spaces, and the representations of minimal fundamental weights
of simply laced Lie algebras. Namely, among all K\"ahler coset spaces
of the form $G/H$, those which are hermitian symmetric and for which is
$G$ is simple and simply laced are distinguished by the fact that the
cohomology ring of $G/H$ is integrable. In other words, there is a
potential $W_{G/H}$ such that
\eqn\cohoms{H^* (G/H) \cong \CJ (W_{G/H})}
These Landau-Ginzburg models are known from \refs{\LVW,\LercheW} as SLOHSS models
(the O stands for ``level one'' and refers to the level of the associated
Kazama-Suzuki model which describes the conformal fixed point of the
Landau-Ginzburg theory).

The space $G/H$ is compact hermitian symmetric precisely if $H$ is a regular,
diagram subalgebra coming from deleting a node of the Dynkin diagram of
$G$ with dual Coxeter number equal to one. The weight $\Xi$ of $G$ with Dynkin
label $1$ on this deleted node and zero elsewhere is a so-called minimal
fundamental weight, and as it turns out, the grading on the cohomologies
\cohoms\ is precisely such that the graded dimension of $\CH$ gives the
polynomial invariant of the unknot associated with the representation
$R$ with highest weight $\Xi$. More precisely \LVW,
\eqn\unnum{{\rm gdim}(\CH) = q^{(\rho,\Xi)} \chi_\Xi (q^\rho)}
where $\rho$ is the Weyl vector of $G$ and $\chi_\Xi$ is the character of
the representation, $R$. Up to the shift in grading by $(\rho,\Xi)$ this
is nothing but the unknot invariant associated with $R$.

Besides the Grassmannians
\eqn\grasco{{\it Gr}(k,N) = {SU(N) \over SU(k)\times SU(N-k)\times U(1)}}
associated with the $k$-th anti-symmetric representation of $SU(N)$,
and the cosets \socoset, associated with the vector representation of $SO(N)$,
there is another series of SLOHSS models built on the classical Lie
groups, associated with the spinor representations of $so(N)$ (for $N$
even),
\eqn\socosets{{ SO(N) \over U(N/2)}}

The cohomology of the coset \socosets\ has generators $z_{2i-1}$ in degree
$2i-1$ for $i=1,2,\ldots, [{N\over 4}]$ and relations which are integrable
to a potential of total degree $N-1$. Although there is no closed general
form for this potential, there is a straightforward algorithm which allows
computation of the potential for any given $N$. The relations are simplest
to see by viewing the coset \socosets\ as the space of complex structures
on $\IR^{N}$. As for the Grassmannians, the cohomology is generated by the
Chern classes of the tautological bundle $E$ in the exact sequence
\eqn\tautological{E\to \IC^{N} \to E^*}
Expanding
$$
c(E)=1+\sum_{i=1}^{N/2} t^i z_i \qquad
c(E^*) = 1+\sum_{i=1}^{N/2} (-1)^i t^i z_i \,,
$$
the relations amongst the $z_i$ can be obtained from the equation
$$
c(E)\cdot c(E^*) =1
$$
The variables $z_{2i}$ for $i=1,\ldots, N/2$ can be immediately eliminated,
leaving us with relations in degree $N-2i$ for $z_{2i-1}$ for
$i=1,2\ldots, [{N\over 4}]$.

\example{$so(10)$} In this case, there are two variables $z_1$ and $z_3$
in degree $1$ and $3$, respectively, and the potential looks like
$$
W_{so(10),S}= {5\over 576} z_1^9 -{1\over 8}{z_1^6z_3}+{1\over 2}
z_1^2z_3^2 -{1\over 3}z_3^3
$$
%$$
%W_{so(10),S} = {2372 \over 140625} x^9 + {92 \over 625} x^6 y +
%{36 \over 25} x^3 y^2 + {1 \over 6} y^3
%$$
\endexample

%%%%%%%%%%%%%%%%%%%%%%%%%%%%%%%%%%%%%%%%%%%%%%%%%%%%%%%%%%%%%%%%%%%%%%%

\medskip\noindent{{\it `Exceptional' Knot Homology}}

Finally, there are two 'exceptional' cosets,
\eqn\eecosets{ { E_6 \over SO(10) \times U(1)}
\quad\quad,\quad\quad {E_7 \over E_6 \times U(1)} }
corresponding to the $27$ and $56$-dimensional representations of $E_6$
and $E_7$, respectively. The superpotentials can be computed by the
same methods as above, using a tautological exact sequence similar to
\tautological. This was done explicitly in \LercheW, and the result
is (the indices on the variables indicate their degrees):
\eqn\wesix{ W_{E_6,27} = z_1^{13} - {25 \over 169} z_1 z_4^3 + z_4 z_1^9 }
%
%%
%\eqn\weseven{ W_{E_7,56} = {100476 \over 19} \Big( x^{19} + x z^2
%+ y^2 z + 37 \Big( {19 \over 2791} \Big)^{3/4} x^{14} y - 21 \Big(
%{19 \over 2791} \Big)^{1/2} x^{10} z \Big) }
%%
%
\eqn\weseven{ W_{E_7,56} =
{2791\over 19} z_1^{19} + 37 z_1^{14} z_5 - 21 z_1^{10} z_9 + z_5^2 z_9 +
    z_1 z_9^2 }
%

%%%%%%%%%%%%%%%%%%%%%%%%%%%%%%%%%%%%%%%%%%%%%%%%%%%%%%%%%%%%%%%%%%%%

\subsec{Totally Symmetric Representations}

In this subsection, we will derive Landau-Ginzburg potentials corresponding
to the totally symmetric representations of $sl(N)$. These representations
do not correspond to minimal fundamental weights, and there is (as far as we
know) no interpretation in terms of the cohomology of some homogeneous space.
To the best of our knowledge, the potentials of this subsection are new.

We recall the generating function of the potentials \Wanti\ for the
cohomology of the Grassmannians $Gr(k,N)$ \refs{\GepnerGR,\Wittengrass}. The
tautological sequence
$$
E \to \IC^N \to F
$$
says $c(E)\cdot c(F)=1$. The cohomology ring is generated by the
cohomology classes of $E$, $c(E)=1+\sum_{i=1}^k t^i z_i$, and the
relations are
\eqn\firstrels{
R_{N+1-i} = 0 \qquad {\rm for }\;\; i=1,\ldots, k\,,}
where the $R_i$ are defined by
$$
c(F) = c(E)^{-1} = 1+\sum_{i\ge 1} t^i R_i (z_i)
$$
It is easy to see that if we define $W_{sl(N),\Lambda^k}$ by the generating
function
\eqn\generate{
\sum_{N=1}^\infty (-1)^{N}t^{N+1} W_{sl(N),\Lambda^k} (z_i) = \log \bigl(1+\sum_i t^i z_i\bigr)
}
then
$$
{\partial W_{sl(N),\Lambda^k} \over \partial z_i} = R_{N+1-i} (z_i)
$$
Introducing the roots $x_i$ of the Chern polynomial
$c(E) = \prod_{i=1}^k (1+t x_i)$, we obtain the form \Wanti\
for $W_{sl(N),\Lambda^k}$.

There is an alternative derivation of $W_{sl(N),\Lambda^k}$ which is closer in 
spirit to Landau-Ginzburg theory and which will be our approach to derive the
potentials for the totally symmetric representations. Let us first
illustrate this in the simplest case $k=2$.

%%%%%%%%%%%%%%%%%%%%%%%%%%%%%%%%%%%%%%%%%%%%%%%%%%%%%%%%%%%%%%%%%%

\medskip\noindent
{\it Rank 2:}

The derivation of the potentials is based on the tensor product
decomposition of irreducible representations of $sl(N)$,
\eqn\decomp{ V \otimes V = \Lambda^2 \oplus S^2 }
As before, $V$ is the vector representation, and $\Lambda^2$
and $S^2$ are the rank 2 anti-symmetric and symmetric
representations of $sl(N)$, respectively.

We can understand the decomposition \decomp\ at the level of
the cohomology by looking at the ground states of the
Landau-Ginzburg theory. In the tensor product, with superpotential
$W_{N\otimes N}=x_1^{N+1}+ x_2^{N+1}$, those ground states can be
obtained by acting with the elements of the chiral ring $\CH^*(N\otimes
N)\cong \CJ_{N\otimes N}=\IC[x_1,x_2]/\langle x_1^N,x_2^N\rangle$ on the
unique groundstate with lowest R-charge $|0\rangle=|0\rangle_1\otimes
|0\rangle_2$. This ground state is clearly symmetric with respect
to exchange of $x_1$ and $x_2$. Acting with only the symmetric combinations
$z=x_1+x_2$ and $w=x_1x_2$, we obtain all symmetric ground states.
Similarly, we obtain all anti-symmetric ground states by acting with
$z$ and $w$ on the lowest one, $|x_1-x_2\rangle = |x_1\rangle_1\otimes
|0\rangle_2-|0\rangle_1\otimes |x_2\rangle_2$.

The relations among $z$ and $w$ are obtained by restricting the tensor
product relations $x_1^N\equiv 0$, $x_2^N\equiv 0$ to the symmetric/anti-symmetric
sector. In the anti-symmetric sector, the relations are generated by $x_1^N-
x_2^N\equiv 0$, $x_1^{N+1}-x_2^{N+1}\equiv 0$, which in terms of $z$
and $w$ become
\eqn\rels{\eqalign{{x_1^N-x_2^N\over x_1-x_2} =
x_1^{N-1} + x_1^{N-2} x_2 + \cdots + x_2^{N-1} & \equiv 0 \cr
{x_1^{N+1}-x_2^{N+1}\over x_1-x_2} =
x_1^{N} + x_1^{N-1} x_2 + \cdots + x_2^{N} & \equiv 0 }}
It is easy to check that these are precisely the relations deduced from
the potentials $W_{sl(N),\Lambda^2}$ with generating function
\eqn\gen{\sum_{N} (-1)^N t^{N+1}W_{sl(N),\Lambda^2}(z,w)=\log(1+tz+t^2w)}

To get the symmetric cohomology, we concentrate on the symmetric
ground states of the tensor product theory. The relations are
simply $x_1^N+x_2^N\equiv 0$, $x_1^{N+1}+x_2^{N+1}\equiv 0$,
which we can write in terms of $z$ and $w$ as
\eqn\srels{ W_{sl(N-1),\Lambda^2}(z,w) \equiv 0 \qquad
\qquad W_{sl(N),\Lambda^2}(z,w) \equiv 0 }
These relations can be integrated to a potential $\tilde W_{sl(N),S^2}$
which is encoded in the generating function
\eqn\final{ \sum (-1)^{N} t^{N+2} \tilde W_{sl(N),S^2} (z,w) =
(1+t z + t^2 w) \log(1+t z + t^2 w) }
Indeed, by taking derivatives with respect to $z$ and $w$
on both sides, and using \gen, we find
$\p_z \tilde W_{sl(N),S^2} = W_{sl(N),\Lambda^2}$ and
$\p_w \tilde W_{sl(N),S^2} =  - W_{sl(N-1),\Lambda^2}$.

%%%%%%%%%%%%%%%%%%%%%%%%%%%%%%%%%%%%%%%%%%%%%%%%%%%%%%%%%%%%%%%%%%

\medskip\noindent
{\it Higher Symmetric Representations}

The foregoing has an immediate generalization to all totally
symmetric/anti-symmetric representations, obtained as subsectors of
the tensor product theory with potential $W_{N^{\otimes k}} = \sum
x_i^{N+1}$. The $k$-th anti-symmetric representation is obtained by
acting with the elementary symmetric functions on the anti-symmetric
state
$$
|\Delta\rangle = \prod_{i<j} (x_i-x_j) |0\rangle
$$
where $\Delta = \det_{r,s} x_r^{k-s}$ stands for the Vandermonde
determinant. The relations of the chiral ring $\CH^*(N^{\otimes
k})=\IC[x_1,\ldots,x_k]/\langle x_1^N,\ldots x_k^N\rangle$, when
restricted to the anti-symmetric sector, are generated for $i=1,2,\ldots,
k$ by $\det_{r,s} x_r^{\la_s+k-s}\equiv 0$, where $\la_s=N+1-i$ for $s=1$
and $0$ else. For the $z_i$'s, this can be written in terms of the
Schur polynomials
$$
S_{N+1-i} = {\det_{r,s} x_r^{\la_s+k-s}\over \Delta} \equiv 0
$$
By a well-known formula (Giambelli's formula), $S_{i}$ is also equal to
the coefficient of $t^i$ in the expansion of $1/\prod (1+t x_j)$ (see, eg,
\fulton). This is what we have been calling $R_i$ in \firstrels, thus 
completing the derivation of the potential for the anti-symmetric representation.

In the symmetric sector, the relations are generated by
\eqn\tot{\eqalign {x_1^N+\cdots + x_k^N &= 0 \cr
x_1^{N+1} + \cdots + x_k^{N+1} &= 0 \cr
\vdots \quad\qquad &   \cr
x_1^{N+k-1} + \cdots + x_k^{N+k-1} &=0
}}
When expressed in terms of $z_1,\ldots z_k$, the
relations are $W_{sl(N+k-i-1),\Lambda^k} =0$ for $i=1,\ldots k$.
Those relations can be integrated,
$W_{sl(N+k-i-1), \Lambda^k}=\p_{z_i} \tilde W_{sl(N), S^k}$
where the $\tilde W$'s are encoded in the generating function
\eqn\gen{ \sum t^{N+k} (-1)^N \tilde W_{sl(N), S^k} (z_1,\ldots,z_k)
= \bigl(1+ \sum_{i=1}^k t^i z_i\bigr) \log\bigl(1+\sum_{i=1}^k t^i z_i \bigr)}
%

%%%%%%%%%%%%%%%%%%%%%%%%%%%%%%%%%%%%%%%%%%%%%%%%%%%%%%%%%%%%%%%%%%%%%%%%%55

\subsec{Representations without Potentials}

Finally, let us give an example of a representation for which the
relations in the homology ring of the unknot cannot be integrated
to a potential. We consider the tensor product decomposition of
three fundamentals
$$
\tableau{1}^{\otimes 3} = \tableau{1 1 1} \oplus 2 \tableau{2 1}
\oplus \tableau{3}
$$
We know already that by acting with the symmetric functions on the
ground state, we obtain the symmetric representation $\tableau{3}$,
while acting on the totally anti-symmetric state $(x_1-x_2)(x_1-
x_3)(x_2-x_3)$, we obtain the representation $\tableau{1 1 1}$.
The remaining states must comprise two copies of the representation
$\tableau{2 1}$. In degree $1$, we have (modulo the symmetric
ones) the two states
$$
|x_1-x_2\rangle \qquad\qquad |x_2-x_3\rangle
$$
We can identify those with the highest weight vectors and build the
representation by acting with polynomials of a certain symmetry type.
But the resulting relations are not integrable to a potential.

For instance, if we act only with the totally symmetric
polynomials, we obtain one states in degree $2$ from each of $x_1-x_2$
and $x_2-x_3$. Since the total number of states in degree $2$ is $6$, and
two are contained in $\tableau{1 1 1}$, we would be missing $2$ states.

On the other hand, if we act for example on $|x_1-x_2\rangle$
with polynomials that are only symmetric with respect to $1\leftrightarrow 2$,
namely $x_1+x_2$, $x_3$ and $x_1 x_2$. Then we get just the right
number of states in degree $2$. But in degree $3$, we will also
get the totally anti-symmetric state $(x_1-x_2)(x_1-x_3)(x_2-x_3)$.
This means that we have an additional relation between $x_1+x_2$,
$x_3$ and $x_1 x_2$ in degree $2$, together with some other
relations in higher degrees. These relations cannot be integrated
to a potential.

%%%%%%%%%%%%%%%%%%%%%%%%%%%%%%%%%%%%%%%%%%%%%%%%%%%%%%%%%%%%%%%%%%%%

\newsec{Deformations}

Given an effective Landau-Ginzburg description, it is natural to
ask for the behavior of the theory under deformations. In the
present context, we could in principle study the dependence of the 
knot homologies on deformations of the matrix factorizations used
to defining them, or on the deformation of the potentials. In
general, we will need to deform both.

Although we know the requisite matrix factorizations only in a limited
number of cases, we can obtain useful information just from the deformation
of the potential associated with a single crossing-less strand with
a single marking,
\eqn\deform{W \leadsto W + \Delta W}
where $\Delta W$ is a polynomial
in the same variables as $W$. In those cases in which the matrix
factorizations are known, we can also study how they must change 
under \deform, and this will give us further valuable information as
well.\foot{In general, we cannot exclude the possibility that after the
deformation, the factorizations lose some of their important properties
for the definition of knot homologies. It could also happen that the
factorization cannot be deformed together with the potential, thereby 
obstructing the deformation altogether. By experience, however, such 
phenomena are rare, and we will ignore them.  See \refs{\HoriJW,\Emanuelii}
for some aspects of the deformation theory of matrix factorizations.}

As we have reviewed above, the definition of Khovanov-Rozansky homology
involves, in physical terminology, vector spaces of open strings between 
D-branes which are described by matrix factorizations of Landau-Ginzburg
potentials. These vector spaces are endowed with a multiplication structure
by associating linear maps with cobordisms of knots (see \RKhovanov\ for 
details). This defines a ring structure on the homology of the unknot, and 
the homology of a general knot becomes a module over $\CH({\rm unknot})$.
On general grounds, we are expecting that under a marginal 
deformation (formally, a deformation respecting the 
grading, $\deg\Delta W=\deg W$), the vector spaces themselves will not 
change, and only the ring/module structure associated with cobordisms 
of knots will be deformed in a certain way. On the other hand, under 
relevant deformations (namely, those with $\deg\Delta W< \deg W$), we 
are expecting that the dimension of the vector spaces will change 
(more precisely, it should decrease).

The purpose of this section is to describe some of these deformations,
and their expected relation to structural elements of the knot homology 
theories.

%%%%%%%%%%%%%%%%%%%%%%%%%%%%%%%%%%%%%%%%%%%%%%%%%%%%%

\subsec{Deformations $sl(N)\leadsto sl(M)$}

Deformations of homological knot invariants were first 
studied by Lee \ESLii\ in the case of Khovanov homology, and further
considered by Bar-Natan \DBNi\ and Turner \turner. See also \Khovanovv\
for further explanations. In \gornik, Gornik considers a deformation of 
Khovanov-Rozansky's $sl(N)$ homology, in which the potential $x^{N+1}$ 
associated with a thin edge is deformed to $x^{N+1} + \beta^{N} x$ (where 
$\beta$ is a scalar parameter), but all the other essentials of the definitions 
of \RKhovanov\ are left unchanged. Because the deformed potential is not 
homogeneous, one thereby obtains instead of a bigraded complex a filtered 
chain complex $\CC^{\rm def}(K)$ for any given knot $K$. Gornik 
concentrates on the unreduced version of the theory and proves two 
statements about $\CC^{\rm def}(K)$ (similar statements hold for Lee's 
deformation of Khovanov homology). The second term in the spectral 
sequence associated with $\CC^{\rm def}(K)$ is isomorphic to the 
undeformed $sl(N)$ homology of Khovanov-Rozansky. Moreover, the 
cohomology of $\CC^{\rm def}(K)$ is, for any knot $K$, and for any $N$, 
isomorphic to the $sl(N)$-homology of the unknot. In the reduced theory, 
the cohomology of the deformed complex is expected to be one-dimensional.

In ref.\ \DGR, it was proposed that the deformed theories of Lee and Gornik
could be usefully mounted into the triply-graded homology theory $\CH^{\rm 
HOMFLY}$ categorifying the (normalized) HOMFLY polynomial. Recall that this 
theory is expected to come equipped with a family of anti-commuting differentials 
$\{d_N\}$, with $N\in\IZ$. The cohomology of $\CH^{\rm HOMFLY}$ with respect 
to $d_N$ with $N>1$ is isomorphic to the (reduced) $sl(N)$ homology 
$\CH^{sl(N)}$ of Khovanov-Rozansky. Differentials $d_{\pm 1}$ are ``canceling'', 
in the sense that their cohomology is one-dimensional in a particular degree. 
By restricting $d_{\pm 1}$ to $\CH^{sl(N)}$, one induces canceling 
differentials on Khovanov-Rozansky homology. The existence of precisely
such a differential follows from the work of Lee and Turner on $sl(2)$ and 
can be derived from Gornik's deformation in the general case \DGR.

\ifig\resolution{Behavior of critical points under the deformation 
(4.2) ($N-M=3$).}
{\epsfxsize4.0in\epsfbox{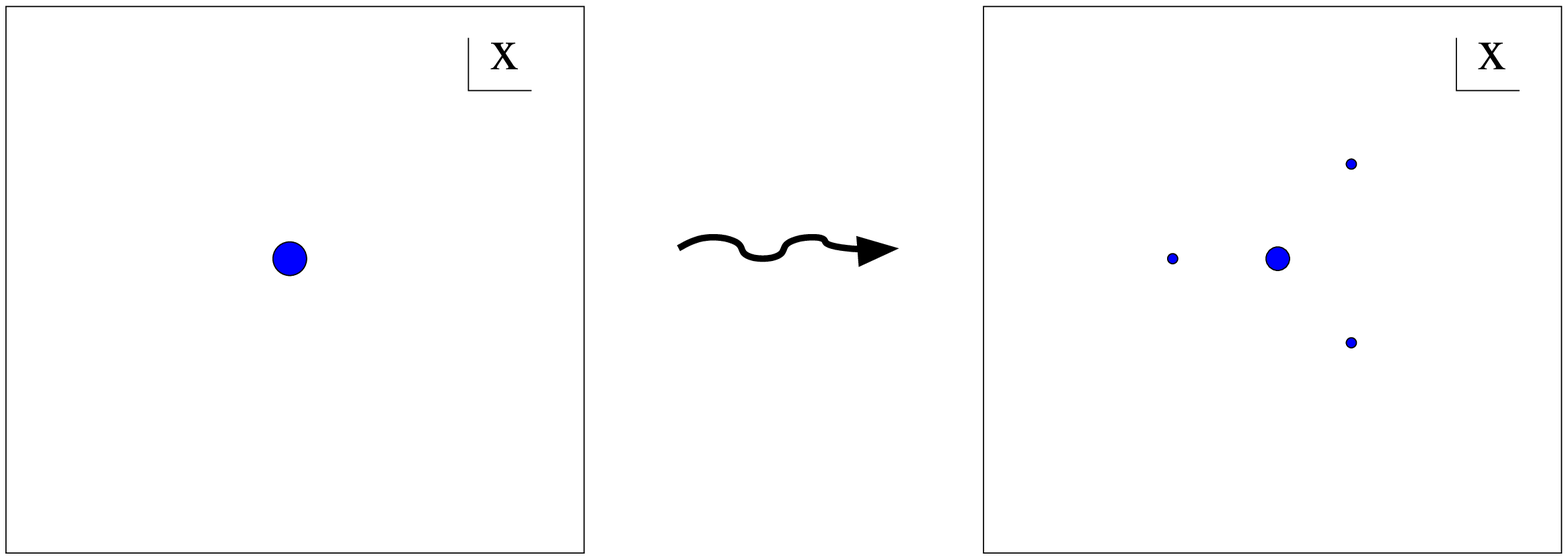}}

More generally, we can consider deformations of the $sl(N)$ potential 
$W_{sl(N)}=x^{N+1}$ by any monomial of lower degree,
\eqn\potdef{ W = x^{N+1}\leadsto x^{N+1} + \beta^{N-M} x^{M+1} }
Under this deformation, the critical point of order $N$ at the origin
$x=0$ is resolved into $N-M$ non-degenerate critical points at $x^{N-M}
=-(M+1)/(N+1)$ and a degenerate critical point of order $M$ at the origin,
which is equivalent to the potential $W_{sl(M)}=x^{M+1}$ (see \resolution). 
One therefore expects that the deformed theory will be related to the 
$sl(M)$ theory, and at the reduced level, give rise to a differential 
$d_{N\to M}$, identified with the restriction of $d_M$ to $\CH^{sl(N)}$.

Let us make this deformation of Khovanov-Rozansky theory more explicit.
As in \gornik, we keep the planar graph calculus intact. In a given planar 
graph, we assign the factorization
\eqn\factdef{
x^{N+1} + \beta^{N-M} x^{M+1}
- y^{N+1} - \beta^{N-M} y^{M+1}
= (x-y) \pi(x,y) = (x-y) (\pi_N(x,y) + \pi_M(x,y))
}
to a single arc (oriented edge) with endpoints labeled $x$ and $y$. Here,
\eqn\arcdef{
\pi_N(x,y) = x^{N} + x^{N-1} y + \cdots + y^{N}}
Now recall that to a wide edge with four thin edges attached
carrying labels $x_1$, $x_2$, $x_3$, $x_4$, Khovanov-Rozansky
associate the factorization
\eqn\thickdef{
x_1^{N+1} + x_2^{N+1} - x_3^{N+1}-x_4^{N+1}
= (x_1+x_2-x_3-x_4) u_{1,N}
+ (x_1 x_2-x_3 x_4) u_{2,N}
}
where
\def\frac#1#2{{#1\over#2}}
\eqn\udef{
\eqalign{
u_{1,N} &= \frac{x_1^{N+1}+x_2^{N+1}-g(x_3+x_4,x_1x_2)}{x_1+x_2-x_3-x_4} \cr
u_{2,N} &= \frac{g(x_3+x_4,x_1x_2)- x_3^{N+1}-x_4^{N+1}}{x_1x_2-x_3x_4}}
}
In the deformed case, we simply write
\eqn\simpledef{
\sum_{i=1}^4 \pm (x_i^{N+1} + \beta^{N-M} x_i^{M+1})
= (x_1+x_2-x_3-x_4) u_{1}
+ (x_1 x_2-x_3 x_4) u_{2}
}
where $u_i=u_{i,N}+\beta^{N-M} u_{i,M}$.
So, to the two resolutions of a crossing in a planar diagram
of a knot, we have associated matrix factorization
\eqn\plandef{
\eqalign{
Q_1 &= \left[
\pmatrix{ \pi_{14} & x_2-x_3\cr \pi_{23} & x_4-x_1},
\pmatrix{ x_1-x_4  & x_2-x_3 \cr \pi_{23} & -\pi_{14}}
\right],\cr
Q_2 &= \left[
\pmatrix{u_1 & x_1x_2-x_3x_4 \cr u_2 & x_3+x_4-x_1-x_2 },
\pmatrix{x_1+x_2-x_3-x_4 & x_1x_2-x_3x_4 \cr u_2 & -u_1}
\right]}
}
The maps $\chi_0$ and $\chi_1$ used by Khovanov-Rozansky
have the form (this corresponds to setting $\mu=0$ and $\lambda=1$
in the definitions of \RKhovanov.)
\eqn\mapdef{
\eqalign{
\chi_0 &= \left[
\pmatrix{ x_4-x_2 & 0 \cr a_1 & 1 } ,
\pmatrix{ x_4 & -x_2 \cr -1 & 1 }\right],
\cr
\chi_1 &= \left[
\pmatrix{1 & 0 \cr -a_1 & x_4-x_2 },
\pmatrix{1 & x_2 \cr 1 & x_4 }\right]
}
}
$\chi_0$/$\chi_1$ is in the cohomology ${\rm Hom}(Q_1,Q_2)$/%
${\rm Hom}(Q_2,Q_1)$ if
\eqn\cohodef{
\eqalign{
\pmatrix{u_1 & x_1x_2-x_3x_4 \cr u_2 & x_3+x_4-x_1-x_2 }
\pmatrix{ x_4-x_2 & 0 \cr a_1 & 1 }
&=
\pmatrix{ x_4 & -x_2 \cr -1 & 1 }
\pmatrix{ \pi_{14} & x_2-x_3\cr \pi_{23} & x_4-x_1} \cr
\pmatrix{1 & 0 \cr -a_1 & x_4-x_2 }
\pmatrix{x_1+x_2-x_3-x_4 & x_1x_2-x_3x_4 \cr u_2 & -u_1}
&=
\pmatrix{ x_1-x_4  & x_2-x_3 \cr \pi_{23} & -\pi_{14} }
\pmatrix{1 & x_2 \cr 1 & x_4 }
}
}
All these equations are solved by
\eqn\solvdef{
a_1=-u_2+\frac{u_1+x_1u_2-\pi_{23}}{x_1-x_4}
}
and it is clear that this works in the deformed case as well.
Note that the property $\chi_0\chi_1=(x_4-x_2)$ is also preserved.

So we can follow all the steps in ref.\ \RKhovanov\ to define a
deformed $sl(N)$ homology of links. Let us denote it by
$\CH_{N\to M}$. What are its properties?

First of all, as in Gornik's case, the deformed differential is a sum
of two terms of different degree of homogeneity, so that we obtain a 
filtered chain complex instead of a bigraded one. The proof of ``Theorem 1'' 
of \gornik\ can then be extended without much difficulty to show the 
analogous statement for the present deformation. Namely, the second term 
in the spectral sequence associated with the filtered chain complex is 
isomorphic to the undeformed $sl(N)$ homology. 

%%%%%%%%%%%%%%%%%%%%%%%%%%%%%%%%%%%%%%%%%%%%%%%%%%%%%%%%%%%%%%%%%%%%%%%%%%%

Gornik's second result, concerning the cohomology of the deformed complex,
is slightly more complicated to generalize. In the unreduced case,
we are expecting that the $sl(M)$ homology will be contained as a summand
in the deformed cohomology, with remaining pieces being $N-M$ dimensional,
independent of the knot.
\eqn\deform{
\CH_{N\to M}(K) \cong \CH_M(K) \oplus \IC^{N-M}
}
Intuitively, the extra terms are associated with the non-degenerate critical
points of the deformed potential (see \resolution). Consequently, going to 
reduced theory will remove these extra terms.

Thus, we see that there is a correspondence between 
the differentials $d_N$ of the triply graded HOMFLY theory, and the 
relevant deformations of the Landau-Ginzburg potentials $W_{sl(N)}=
x^{N+1}$. The degree of the differentials can also be (partially)
understood from this interpretation: When restricted to $\CH^{sl(N)}$, 
$d_M$ has $q$-degree $2(M-N)$ (see \DGR), which matches the ``relevance'' 
of the deformation,
\eqn\degreematch{\deg x^{M+1}- \deg x^{N+1} = 2(M-N)}
(recall that $\deg x=2$).

It is natural to generalize these considerations and to look for 
differentials on other homology theories which can be induced from 
relevant deformations of the Landau-Ginzburg potentials of section 3.

%%%%%%%%%%%%%%%%%%%%%%%%%%%%%%%%%%%%%%%%%%%%%%%%%%%%%%%%%%%%%%%%%%%%%%%

\subsec{Deformations $so(N)\leadsto so(M)$ and $so(N)\leadsto sl(N-2)$}

For example, let us consider the $so(N)$ potential $W_{so(N)}=x^{N-1}+x y^2$.
Among the relevant deformations, those with 
a definite grading are the monomials in $x$ as well as the
monomials $y$, $xy$, and $y^2$. We are expecting that each of these 
deformations will give rise to a differential on $\CH^{so(N)}$,
and we can predict the cohomology of these differentials by looking
at the type of singularity of the deformed potential. For instance,
adding $x^{M-1}$ deforms $W_{so(N)}$ into $W_{so(M)}$ so we expect
that the cohomology of the corresponding differential, again denoted
$d_{N\to M}$, will be isomorphic to the $so(M)$ knot homology. 
\eqn\soNtosoM{ (\CH^{so(N)},d_{N\to M}) \cong \CH^{so(M)} }
In particular, for $M=2$, we obtain a canceling differential.

The deformation which we find most interesting for our present purposes
is the one relating the $so$ series of potentials with the $sl$ series.
Consider the deformed potential
\eqn\soNtoslN{ x^{N-1} + x y^2 + y^2}
obtained from adding to $W_{so(N)}$ a quadratic term for $y$. Under
this deformation, the isolated critical point of order $N$ at $x=y=0$ 
is resolved into two non-degenerate critical points at $x=-1$, 
$y^2=(N-1)(-1)^{N-1}$, and a degenerate critical point of order $N-2$ at 
the origin. Around this degenerate critical point, the deformed potential 
is equivalent to the $sl(N-2)$ potential $W_{sl(N-2)}= x^{N-1}$.

We are conjecturing that this deformation can be extended to the
entire $so(N)$ homology theory. The deformation will lead to a 
differential $d_{y^2}$ on the $so(N)$ homology with cohomology 
equivalent to $sl(N-2)$ homology,
\foot{We should mention an important caveat here. It is known from the
context of ``Kn\"orrer periodicity'' that the categories of matrix
factorizations associated with $W(x)$ and $W(x)+y^2$ are {\it not}
strictly equivalent, unless the latter is equivariantized with respect 
to $y\mapsto -y$ \KapustinLii. It remains to be seen how this will affect 
the conjectured relation between $so(N)$ and $sl(N-2)$ homologies.}
\eqn\equivdef{(\CH^{so(N)},d_{y^2})\cong \CH^{sl(N-2)}}
The expected degree of this differential can be determined
by noting that the undeformed potential is homogeneous of degree $2N-2$
if we assign degree $2$ and $N-1$ to $x$ and $y$, respectively.
Thus, $d_{y^2}$ will have degree $-2$, independent of
$N$. We will find evidence for the existence of such a differential
in the triply-graded theory considered in section 6.

We summarize the relevant deformations of $W_{so(N)}$ and associated 
differentials on $\CH^{so(N)}$ in the following table:
\vskip 1em
\centerline{\vbox{\halign{\quad # & \quad # \cr
$\underline{{\rm ~deformed~ potential~}}$ & $\underline{{\rm ~differential~}}$ \cr
$x^{N-1}+x y^2+x^{M-1}$   & $d_{N\to M}\cong d_M|_{\CH^{so(N)}}$ \cr
$x^{N-1}+x y^2 + x$   & canceling \cr
$x^{N-1}+x y^2 + y$   & canceling \cr
$x^{N-1}+x y^2 + x y$ & canceling \cr
$x^{N-1}+x y^2 + y^2$ & $d_{y^2}$ \cr }}}
%\centerline{ \hbox{{\bf Table 1:}
%{\it ~~ Relevant deformations of $W_{so(N)}$ and associated 
%differentials.}}} } \vskip 0.5cm

%Of particular interest are deformations that relate $sl(N)$ and $so(M)$
%homology theories. For example, one can consider the following
%deformation of the potential $W_{so(4)}$:
%$$
%\eqalign{
%W_{so(4)} & = W_{sl(2)} (x) + W_{sl(2)} (y) = x^3+y^3
%\cr & \longrightarrow
%x^3+y^3 + (x+y)^2 = ((x+y)^3 + 3 (x+y) (x-y)^2)/4 + (x+y)^2 }
%$$
%which folws to $W_{so(3)} = z^2 + z w^2$ with $z=x+y$ and $w=x-y$.
%This deformation leads to a differential, $d$, which acting on
%the $so(4)$ knot homology gives $so(3)$ knot homology.
%In other words, there is a spectral sequence which
%starts with $E^2 = \CH^{so(4),V}$ and converges to $E^{\infty} = \CH^{so(3),V}$.
%The differential $d$ must be precisely the differential $d_3$
%that we have in the triply-graded Kauffman homology (see below).

%%%%%%%%%%%%%%%%%%%%%%%%%%%%%%%%%%%%%%%%%%%%%%%%%%%%%%%%%%%%%%%%%%%%%%%

\subsec{Marginal Deformations}

Until now, we have considered only relevant deformations which change
the knot homologies as vector spaces, and relate categorifications
of different type. As we have noted above, marginal deformations are 
expected to change only the algebraic structure on the knot homologies, 
leaving the vector spaces untouched. 

For example, recall from section 3 that the potentials $W_{sl(N),S^k}$
associated with the $k$-th symmetric representation of $sl(N)$ 
are homogeneous functions of total degree $N+k$ in $k$ variables of 
degree $1,2,\ldots,k$. These are precisely the same degrees as for
the potential $W_{sl(N+k-1),\Lambda^k}$ associated with the
$k$-th anti-symmetric representation of $sl(N+k-1)$. Since the
Poincare polynomial of the Landau-Ginzburg model depends only on the
degrees of the variables and the total degree of the potential,
the corresponding homologies are equal as vector spaces,
\eqn\same{\CH^{sl(N),S^k}({\rm unknot}) \cong \CH^{sl(N+k-1),
\Lambda^k}({\rm unknot})}
However, since the potentials are not equivalent by a change of 
variables, the ring structure on the two sides of \same\ will not
be the same.

Based on the existence of such a deformation, it is tempting
to speculate that it can be extended beyond the unknot. It is 
admittedly difficult to make this precise at the moment, given 
that we do not know the combinatorial definition of, e.g., the 
$(sl(N),S^k)$ knot homologies. In fact, the combinatorics will
most likely not be equivalent to the MOY calculus relevant for
the anti-symmetric representations. Hence the corresponding 
categorifications are not expected to be equal at the level of
the knot invariants, and can at most be related at a more subtle 
level. We find it plausible that such a relation 
could exist.

%%%%%%%%%%%%%%%%%%%%%%%%%%%%%%%%%%%%%%%%%%%%%%%%%%%%%%%%%%%%%%%%%%%%%%%

\newsec{Topological Strings and $so(N)$ Knot Homology}

Now let us consider a string theory realization of the polynomial
and homological knot invariants associated with gauge groups
$SO(N)$ and $Sp(N)$. As in the $SU(N)$ case, this will lead us to
important structure theorems for these knot homologies and,
eventually, to a reformulation based on a new triply-graded
theory, that will be discussed in more details in section 6.

\subsec{Embedding in Topological String}

The physical setup for $SO/Sp$ gauge groups can be obtained from
the one for $SU(N)$ by introducing a suitable orientifold projection.
Namely, recall \Wittencsstring, that Chern-Simons gauge theory with a
unitary gauge group can be realized in the topological A-model
by considering D-branes wrapped around $\S^3$ in the deformed
conifold geometry $T^* \S^3$.
This space can be described as a hypersurface in $\IC^4$,
\eqn\defconif{ z_1 z_4 - z_2 z_3 = \mu }
where $\mu$ is a complex deformation parameter.
A theory with $SO(N)$ (resp. $Sp(N)$) gauge group can be obtained
by starting with $N$ D-branes wrapped around $\S^3$ and
introducing an orientifold, which acts on space-time by an
involution
\eqn\definvol{ \tau~:~~ (z_1, z_2, z_3, z_4) \to (\bar z_4, - \bar
z_3, - \bar z_2, \bar z_1)}
More precisely, in order for \definvol\ to be an anti-holomorphic
symmetry of the deformed conifold \defconif, we have to restrict $\mu$
to be real. The fixed point set of the involution (the location of the
$\CO$-plane) is the locus $|z_1|^2+|z_2|^2=\mu$ in $\IC^2$, which for $\mu>0$
is precisely the three-sphere, while for $\mu<0$ it is empty. The minimal
supersymmetric three-cycle in the latter case is $\S^3/\IZ_2\cong\IR\IP^3$.
Wrapping D-branes on the supersymmetric 3-cycle of this orientifolded
conifold leads to a gauge theory on the brane world-volume.
For $\mu>0$ the result is the Chern-Simons gauge theory on $\S^3$
with gauge group $SO(N)$ or $Sp(N)$, depending on the orientifold action
on the Chan-Paton factors, while for $\mu<0$ we obtain
$SU(N)$ Chern-Simons theory on $\IR\IP^3$.

The topological A-model that we are considering here
does  not depend on complex structure deformations such as $\mu$.
We are  therefore led to the conclusion that there is a duality between knot
invariants obtained in $SO(N)/Sp(N)$ gauge theory on $\S^3$ and $SU(N)$
gauge theory on $\IR\IP^3$. In the latter case, the distinction between
the two different orientifold projections corresponds in the gauge theory
to the choice of a discrete Wilson line associated with $H^1(\IR\IP^3)=\IZ_2$.

It is instructive to contrast this expected behavior with the situation in the
physical string, which was recently studied in detail in \HoriBK. First of all,
in that case, one is mostly interested in studying the gauge theory on the
four-dimensional world-volume transverse to the Calabi-Yau. Because of the
reduction to the zero modes, this gauge theory is of $SO/Sp$ type also for
$\mu<0$. Secondly, the charge of the $\CO$-plane on $\S^3$ is non-zero and
is equal to minus twice the charge of a brane wrapped on the 3-cycle.
Therefore, fixing the flux at infinity leads to a jump in the rank of
the gauge group by $2$ when going from $\mu>0$ to $\mu<0$. Finally, and most 
importantly, the transition through $\mu=0$ is expected to be possible 
dynamically only for some special cases of flux and orientifold projection.
Here, we are interested in the topological string on the deformed
conifold with an orientifold and branes, and the story is slightly different.

As in the situation without the orientifold \OV,
one can incorporate Wilson loop observables associated with knots and
links by introducing additional Lagrangian D-branes. For example, given a
knot $K\subset \S^3$, the corresponding Lagrangian submanifold $\CL_K
\subset T^* \S^3$ is defined as a conormal bundle to the knot. In
$SO(N)$ or $Sp(N)$ Chern-Simons theory, we have to further divide
by the involution $\tau$, which reverses the orientation in the
fiber directions of $T^* \S^3$ and acts trivially on the $\S^3$.
Generically, this leads to a singular Lagrangian submanifold
$\CL_K$, which has a conical singularity along the knot.\foot{In
order to avoid the singularity, one can consider a small
deformation of the Lagrangian submanifold $\CL_K$ by moving it
away from the $\S^3$. On the covering space, this corresponds to
considering two copies of the knot (mapped to each other by the
involution $\tau$).}

\ifig\coveringfig{Toric diagram representing a D-brane and its image
on the covering space, the total space of the $\CO (-1) \oplus \CO (-1)$
bundle over ${\bf CP}^1$.}
{\epsfxsize3.0in\epsfbox{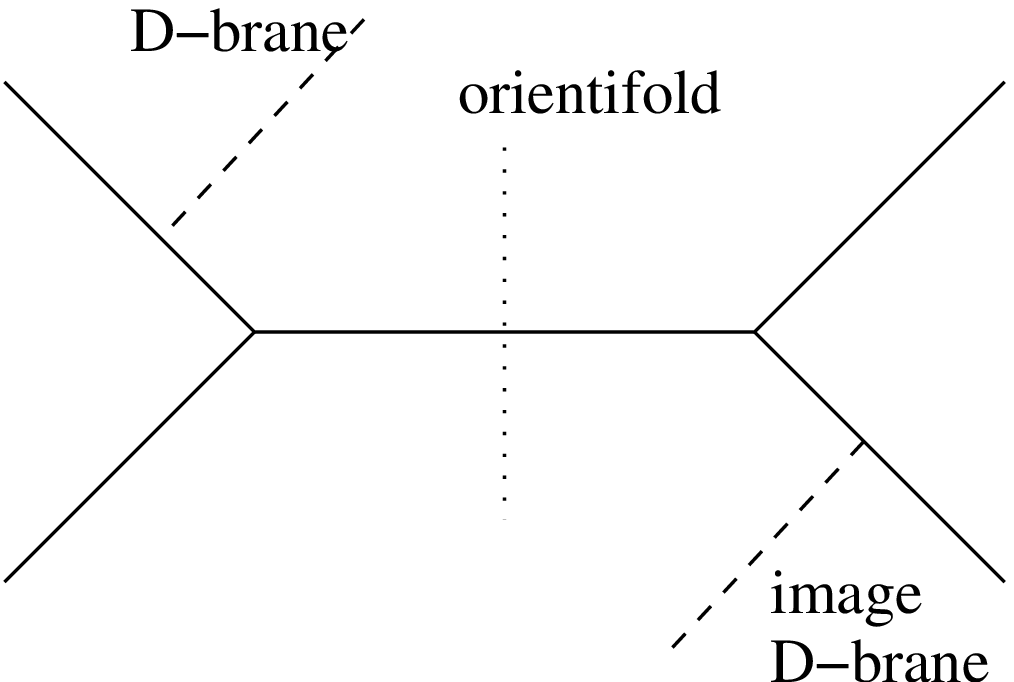}}

In order to see the relation with homological knot invariants, we
need to take $N$ to be large. In this limit, the theory is dual
to topological string theory on the {\it resolved} conifold $X$,
the total space of the $\CO (-1) \oplus \CO (-1)$ bundle over ${\bf CP}^1$.
The space $X$ can be described as a toric variety
(gauged linear sigma model), $X = \IC^4 / \IC^*$,
where $\IC^4$ is parametrized by $X_i$, $i=1,\ldots,4$,
with charges $(1,1,-1,-1)$ with respect to the $\IC^*$ action.
In these variables, the space $X$ is
\eqn\resconif{ X = \{ |X_1|^2 + |X_2|^2 - |X_3|^2 - |X_4|^2 = r \}
/ U(1) }
where $t=r-i\theta$ is the FI parameter (K\"ahler structure modulus).
The value of $t$ determines the volume of the ${\bf CP}^1$ cycle
inside $X$, and is given by the coupling constant and the rank of
the dual Chern-Simons theory, $t = g_s (N+c)$. Instead of $g_s$
and $t$ we shall use the following variables
\eqn\qtvars{\eqalign{ & q = e^{g_s} = \exp \Big( {\pi i \over k +
h} \Big) \cr & \lambda = e^t = q^{N+c} }}
where $h$ is the dual Coxeter number of the gauge group and $c$ is
the charge of the orientifold.

%%%%%%%   TABLE

\vskip 0.8cm \vbox{ \centerline{\vbox{
\hbox{\vbox{\offinterlineskip
%%% usual height for tablespace is 2pt
\def\tablespace{height7pt&\omit&&\omit&&\omit&\cr}
\def\tablerule{\tablespace\noalign{\hrule}\tablespace}

\hrule\halign{&\vrule#&\strut\hskip0.2cm\hfill
#\hfill\hskip0.2cm\cr
\tablespace & Gauge group && $h$ && $c$ &\cr
\tablerule & $U(N)$ && $N$ && $0$ &\cr \tablerule & $SO(N)$ &&
$N-2$ && $~~-1~~$ &\cr \tablerule & $Sp(N)$ && ${N\over 2}+1$
&& $1$ &\cr
\tablespace}\hrule}}}}
\centerline{ \hbox{{\bf Table 1:}{\it ~~ The dual Coxeter number $h$ and 
the orientifold charge $c$.}}} } \vskip 0.5cm

For $SO(N)$ or $Sp(N)$ theories we also need to divide
by the involution $\tau$, which acts on the space $X$ as
\eqn\resinvol{ \tau~:~~ (X_1, X_2, X_3, X_4) \to (\bar X_2, - \bar
X_1, \bar X_4, - \bar X_3)}
In particular, it acts freely on $X$, so that the quotient space
$X/\tau$ contains a 2-cycle ${\bf RP}^2$ instead of ${\bf CP}^1$.

After the large $N$ transition, the branes on $\S^3$ disappear and
we are left only with Lagrangian branes on $\CL_K \subset X$. This
leads to a reformulation of Chern-Simons invariants
%(in particular, quantum group knot invariants)
in terms of open topological string amplitudes on $X$.
The open string amplitudes, in turn, can be expressed
via integer BPS invariants that we discuss next.

%%%%%%%%%%%%%%%%%%%%%%%%%%%%%%%%%%%%%%%%%%%%%%%%%%%%%%

\subsec{BPS States and Homological Knot Invariants}

The open topological string amplitudes on $X$ have certain
integrality properties, which will be important to us below in
understanding homological knot invariants. These properties can be
seen by realizing the setup discussed above in superstring theory.
Namely, following \refs{\OV,\LMV,\FloreaM}, we consider type IIA
string theory on $\R^4 \times X$ together with D4-branes on $\R^2
\times \CL_K$, where $\R^2 \subset \R^4$ and $\CL_K$ is the
Lagrangian submanifold in $X$. For $SO/Sp$ theories, we also need
to introduce an orientifold plane.

In the case of topological strings, we had only two types of
orientifods, $\CO^{\pm}$, which have opposite charge and lead to
$SO$ and $Sp$ gauge groups, respectively. On the other hand, in
the superstring we have orientifolds $\CO^{\pm}$ as well as
anti-orientifolds $\bar \CO^{\pm}$, which lead to same gauge
groups but have opposite Ramond-Ramond charge. In order to
preserve supersymmetry, we must choose D-branes together with
orientifold planes when the total amount of flux, $N$, is
positive, and anti-D-branes together with anti-orientifold planes
$\bar \CO^{\pm}$ when $N<0$. Therefore, in total we have four
different choices summarized in the table below. Note that
the chirality of the knot is also correlated with the sign of
the flux.

%%%%%%%   TABLE -- ORIENTIFOLDS  %%%%%%%%%%%%%%%%%%%%%%%%%%%%%

\vskip 0.8cm \vbox{ \centerline{\vbox{
\hbox{\vbox{\offinterlineskip
%%% usual height for tablespace is 2pt
\def\tablespace{height7pt&\omit&&\omit&&\omit&&\omit&&\omit&\cr}
\def\tablerule{\tablespace\noalign{\hrule}\tablespace}

\hrule\halign{&\vrule#&\strut\hskip0.2cm\hfill
#\hfill\hskip0.2cm\cr
\tablespace & A-model && $N$ && Type IIA theory && Gauge group &&
Knot invariant &\cr \tablerule
& $\CO^-$ && $N>0$ && $\CO^-$ && $SO$ && $F(K;\la=q^N,q)$ &\cr
\tablespace
&  && $N<0$ && $\bar \CO^+$ && $Sp$ && $F(\bar K;\la=q^N,q)$ &\cr
\tablerule
& $\CO^+$ && $N>0$ && $\CO^+$ && $Sp$ && $F(K;\la=-q^N,q)$ &\cr
\tablespace
&  && $N<0$ && $\bar \CO^-$ && $SO$ && $F(\bar K;\la=-q^N,q)$ &
\cr
\tablespace}\hrule}}}}
\centerline{ \hbox{{\bf Table 2:}{\it ~~ Different types of orientifolds in the A-model
and type IIA string theory.}}} } \vskip 0.5cm

\noindent
These four choices correspond to different specializations of the Kauffman
polynomial at $\la = \pm q^N$, where the choice of sign is correlated with
the charge of the orientifold plane.
This interpretation is consistent with eqs. \specializations\ and \negativeN,
\eqn\fcchoices{\eqalign{
& \bar F_{N+1} (q) = \bar F(\la=q^{N},q) = \bar F(\bar K;\la=-q^{-N},q) \cr
& \bar C_{N-1} (q) = \bar F(\la=-q^{N},q) = \bar F(\bar K;\la=q^{-N},q)
}}

Embedding the topological string setup in type IIA string theory
allows to express open string amplitudes and, therefore,
polynomial knot invariants in terms of integer numbers
that count degeneracies of BPS states \refs{\OV,\LMV}.
For example, for the HOMFLY polynomial we have
\eqn\fnhat{\eqalign{ \bar P(\la,q)
& = \sum_{g,Q} \hat N_{g,Q} \la^Q (q^{-1} - q)^{2g-1} \cr
& =  {1 \over q-q^{-1}} \sum_{Q,s} N_{Q,s} \la^Q q^s
}}
where $\hat N_{g,Q}$ (resp. $N_{Q,s}$) denote the degeneracies
of BPS states, counted with $\pm$ signs. Roughly speaking,
these BPS states are represented by genus $g$ holomorphic
Riemann surfaces in $X$ with boundary on the Lagrangian submanifold $\CL_K$.
More precisely, the space of BPS states on the Lagrangian
D4-brane is $\Z_2 \oplus \Z \oplus \Z \oplus \Z$-graded;
it is graded by three integer quantum numbers and the fermion number $F$,
\eqn\bpsspace{ \CH_{BPS} = \CH_{BPS}^{F,Q,s,r} }
Therefore, one can introduce the following ``index''
\eqn\nqsbps{ N_{Q,s} = \sum_{r,F} (-1)^{r+F} \dim \CH_{BPS}^{F,Q,s,r} }
which appears in \fnhat, and its refinement
\eqn\dqsrbps{ D_{Q,s,r} = \sum_F (-1)^F \dim \CH_{BPS}^{F,Q,s,r} }
which is related to the categorification of the HOMFLY polynomial \GSV.

Similarly, in the presence of the orientifold,
there are additional BPS states represented by unoriented
Riemann surfaces. This naturally leads to the well-known
relation between the HOMFLY and the Kauffman polynomial \FloreaM:
\eqn\fvsp{ \bar F(\la,q) - \bar P (\la,q) = \sum_{g,Q}
N^{c=1}_{g,Q} (q-q^{-1})^{2g} \la^Q + \sum_{g,Q}
N^{c=2}_{g,Q} (q-q^{-1})^{2g+1} \la^Q }
where the integer coefficients $N^{c=1}_{g,Q}$ and $N^{c=2}_{g,Q}$
are interpreted as BPS degeneracies associated with
the contribution of one and two crosscaps, respectively.
This relation simplifies further for torus knots since \LPerez,
\eqn\nctwozero{ N^{c=2}_{\tableau{1},g,Q} = 0 }

Extending \nqsbps\ -- \dqsrbps\ to the orientifold case,
we can also write the Kauffman polynomial $\bar F(\la,q)$ for any knot $K$
in terms of the refined BPS invariants $D^{{\rm Kauffman}}_{Q,s,r}$:
\eqn\fviad{
\bar F (\la,q) = {1 \over q-q^{-1}}  \sum_{Q,s,r}  (-1)^r \la^Q q^s D^{{\rm Kauffman}}_{Q,s,r}
}
Notice, that the specialization of this expression to $\la=q^{N-1}$ is very similar
to the categorification of the quantum $so(N)$ invariants, {\it cf.} \fnviah.
In order to make this relation more precise, let us introduce
the graded Poincare polynomials for the $so/sp$ knot homologies
\eqn\funpoincare{\eqalign{
\bHSO_N (q,t) &= \sum_{i,j \in \Z} t^i q^j \dim \CH_{i,j}^{so(N)} (K)  \cr
\bHSp_N (q,t) &= \sum_{i,j \in \Z} t^i q^j \dim \CH_{i,j}^{sp(N)} (K) }}
and similar polynomials, $\HSO_N (q,t)$, $\HSp_N(q,t)$ for the reduced homologies.
Then, comparing \fnviah\ and \fviad\ we naturally arrive to the following
conjecture, which parallels the conjecture of \GSV\ as formulated in \DGR:
\medskip\noindent
{\bf Conjecture 1:} {\it For a knot $K$, there exists a finite
polynomial $\bSK (K) \in \Z [\la^{\pm 1}, q^{\pm 1}, t^{\pm
1}]$ such that}
\eqn\superpol{ \eqalign{
\bHSO_N (K; q,t) &= {1 \over q-q^{-1}} \bSK (K; \la = q^{N-1},q,t) \cr
\bHSp_N (\bar K; q,t) &= {1\over q-q^{-1}} \bSK (K; \la=q^{-N-1},q,t) }}
{\it for all sufficiently large $N$.}

Notice, in order to be consistent with the polynomial
specializations \specializations, the three-variable
polynomial $\bSK (K; \la,q,t)$ must be a refinement of
the Kauffman polynomial, in a sense that
\eqn\superkauff{ \bar F(K;\la,q) = {1\over q-q^{-1}} \bSK (K;\la,q,t=-1) }
In particular, from \fviad\ and \superkauff\ it follows that the
polynomial $\bSK (K; \la,q,t)$ --- which we shall call the
Kauffman superpolynomial below --- is simply the generating
function of the refined BPS invariants,
\eqn\superkauffviad{
\bSK (\la,q,t) = \sum_{Q,s,r} \la^Q q^s t^r D^{{\rm Kauffman}}_{Q,s,r} }

According to Conjecture 1, there should exist many regularities
and relations between $so(N)/sp(N)$ knot homologies. In
particular, in the rest of this section, we demonstrate how this
conjecture can be used to predict the $so(N)/sp(N)$ homological
invariants for all values of $N$. As a starting point, we shall
use the homological $so(N)/sp(N)$ invariants for small values of
the rank $N$, which can be deduced using certain isomorphisms
between $so$, $sp$, and $sl$ Lie algebras of small rank. For
example, using the isomorphism $so(4) \cong sl(2) \times sl(2)$
and the fact that a vector representation of $so(4)$ corresponds
to the representation $({\bf 2},{\bf 2})$ of $sl(2) \times sl(2)$,
we conclude that $so(4)$ homology is isomorphic to the ``square''
of the Khovanov homology:
\eqn\sofourisoma{ \CH^{so(4)} \cong \CH^{sl(2)} \otimes \CH^{sl(2)} }
Note, that this isomorphism holds in both reduced and unreduced
theories. In particular,
\eqn\sfourhso{ \bar{HSO}_4 (q,t) = \bar{Kh} (q,t)^2 }
Similarly, the isomorphism $sp(2) \cong sl(2)$ leads to the
relation between the $sp(2)$ and $sl(2)$ homological invariants:
\eqn\sptwokh{ \bar{HSp}_2 (q,t) = \bar{Kh} (q^2,t) }
The relations \sfourhso\ and \sptwokh\ generalize, respectively,
the relations \sofourpolrel\ and \sptwopolrel\ between the
corresponding polynomial invariants.

\example{The Trefoil Knot}
The unnormalized Kauffman polynomial for the trefoil knot looks like
\eqn\funtrefoil{\eqalign{ \bar F (3_1) = {1 \over q-q^{-1}} \Big[ & - \la
(q^2 + q^{-2}) + \la^2 (q^3 - q^{-3}) + \la^3 (q^2 + 1 + q^{-2}) \cr
& - \la^4 (q^3 - q^{-3}) - \la^5 + \la^6 (q - q^{-1}) \Big]
}}
On the other hand, its $so(4)$ homological invariant can be
obtained from the relation \sfourhso\ with Khovanov homology:
\eqn\hsofourtref{ \bar{HSO}_4 (3_1) = q^2 + 2 q^4 + q^6 + 2 q^6
t^2 + 2 q^8 t^2 + 2 q^{10} t^3 + 2 q^{12} t^3 + q^{10} t^4 + 2
q^{14} t^5 + q^{18} t^6 }
The simplest form of the Kauffman superpolynomial $\bar \CF
(\la,q,t)$ consistent with \funtrefoil\ and \hsofourtref\ is given
by the following expression:
\eqn\trefoilpunred{\eqalign{ \bar \CF (3_1) = & - \la (q^{-2} +
q^2 t^2) + \la^2 (- q^{-3} + q^{-1} - q^{-1} t^2 -q^3 t^3) + \la^3
(q^{-2} + t^2 - t^3 - t^4 + q^2 t^4) \cr & + \la^4 (q^{-3} t^2 + q
t^3 - q t^5 + q^3 t^5) + \la^5 (q^{-2} t^3 + t^5 - q^{-2} t^5) +
\la^6 (- q^{-1} t^6 + q t^6) }}
Indeed, it is easy to verify that specializations of this
expression to $\la=q^3$ and $t=-1$ yield, respectively, the
homological $so(4)$ invariant and the Kauffman polynomial, in
agreement with \superpol\ and \superkauff.
\endexample

This example can easily be generalized to all torus knots of type $(2,2k+1)$:
\eqn\funredtorusknots{ \eqalign{\bar\CF(T_{2k+1,2}) =
 & (\lambda-\lambda^{-1} + q-q^{-1}) (\lambda/q)^{2k} \cr
& + (1+\la q t)\la^{2k+1}\Bigl[
 (1-q^2\la^{-2})(1+\la q^{-3})\sum_{i=1}^k t^{2i}q^{4i-2k-2}  \cr
& + (1-q^{-2})(1+\la^2
q^{-2}t)\sum_{j=0}^{k-1}\;\sum_{i=0}^{2k-2j-2} t^{2i+2j+4}\la^i
q^{i+4j-2k+4}\Bigr]
%& + \lambda^{2k+1}(1+\lambda q t)(1-q^2\lambda^{-2})(1+\lambda q^{-3})
%\sum_{i=1}^k t^{2i} q^{4i-2k-2} \cr
%& + \lambda^{2k+1}(1+\lambda q t)(1-q^{-2})(1+\lambda^2 q^{-2}t)
%\sum_{i=0}^{2k-2} t^{2i+4} \lambda^i q^{i-2k+4}
%\sum_{j=0}^{{\rm min}(i,2k-i-2)} q^{3j} \lambda^{-j}
}}
Several comments are in order regarding the structure of this
expression. First, as prescribed by \superpol\ and \superkauff, it
reduces to the square of the Khovanov homology at $\lambda=q^3$,
and to the Kauffman polynomial at $t=-1$. The first term in
\funredtorusknots\ corresponds to the homology of the unknot,
while both the second and the third term correspond to parts in
the homology that can be ``killed'' by the differential of the
appropriate degree. Moreover, the structure of the second term is
very similar to the result eq.\ (89) in \DGR\ for the unreduced
superpolynomial $\bar\CP$ of $T_{2k+1,2}$. Also, the last term in 
\funredtorusknots\ is multiplied by a factor $(q-q^{-1})$. This 
structure suggests that, in the topological string interpretation, 
the second term comes from the oriented worldsheets and the third 
term from the unoriented ones, {\it cf.} \fvsp.

Let us consider a more complicated example:

\example{The Figure-eight Knot}
The unnormalized Kauffman polynomial for the figure-eight knot is
\eqn\funfeight{\eqalign{ \bar F(4_1) =
{1 \over q-q^{-1}} \Big[ & \la^{-3}(1-q^2-q^{-2})+\la^{-1}(q^4+q^{-4})
+ q-q^{-1} - \la(q^4+q^{-4})   \cr & + \la^3(-1+q^2+q^{-2}) \Big]
}}
Using constraints from various specializations we are led to the following
prediction for unreduced Kauffman superpolynomial:
\eqn\figeight{
\eqalign {\bar \CF (4_1) = &
- \la^{-3} q^{-2} t^{-4} + \la^{-3} t^{-4}
- \la^{-3} q^2 t^{-2} - \la^{-2} q^{-1} (t^{-2} + t^{-3})
- \la^{-1} q^{-4} t^{-3} \cr &
+ \la^{-2} q (t^{-3}-t^{-1}) + \la^{-1} q^{-2}(t^{-3}-t^{-1})
+ \la^{-1}(-1+t^{-2})
+ \la^{-2} q^5(-1-t) \cr &+ \la^{-1} q^2 (-1-t) - q^{-1}
- \la q^{-4} + \la^2 q^{-7} (-1+t^{-2})
+\la^{-2}q^7(1-t^2) \cr & + \la^{-1}q^4 + q + \la q^{-2}(1+t^{-1})
+\la^2q^{-5}(1+t^{-1}) + \la(1-t^2) + \la q^2(t-t^3)\cr & +\la^2q^{-1}(t-t^3)
+\la q^4t^3+\la^2q(t^2+t^3)+\la^3 q^{-2}t^2
-\la^3t^4+\la^3q^2t^4}}
It reduces to the Kauffman polynomial at $t=-1$ and to the
Poincare polynomial of the $so(4)$ knot homology, $\bar{HSO_4}
(4_1) = \bar{Kh} (4_1)^2$, at $\la=q^3$.
\endexample

%%%%%%%%%%%%%%%%%%%%%%%%%%%%%%%%%%%%%%%%%%%%%%%%%%%%%%

\subsec{Comparison with Colored Khovanov Homology}

Another useful relation follows from the isomorphism $so(3) \cong
sl(2)$. Namely, since the vector representation of $so(3)$
corresponds to a 3-dimensional (spin-1) representation of $sl(2)$,
we have an isomorphism
\eqn\sothreeisom{ \CH^{so(3)} (K) \cong \CH^{sl(2);V_3} (K) }
where we use the notation $\CH^{sl(2);V_n} (K)$ for the $sl(2)$
homology of the knot $K$ colored by the $n$-dimensional
representation $V_n$. This theory should provide a
categorification of the colored Jones polynomial,
\eqn\jneuler{ \bar J_n (q) = \sum_{i,j} (-1)^i q^j \dim
\CH_{i,j}^{sl(2);V_n} (K)}
Recall, that the colored Jones polynomial $\bar J_n (K)$ can be
expressed in terms of the ordinary Jones polynomial of the cables
of the knot $K$:
\eqn\jnviajones{ \bar J_n (K) = \sum_{k=0}^{[{n-1 \over 2}]} (-1)^k
{n-k-1 \choose k} \bar J (K^{n-2k-1}) }
Recently, Khovanov \Khovanovii\ proposed a similar definition of
the homology $\CH^{sl(2);V_n} (K)$ in terms of the ordinary
$sl(2)$ homology of the cables of $K$. For example, the
homology $\CH^{sl(2);V_3} (K)$ associated with the
three-dimensional representation of $sl(2)$ fits into the
following long exact sequence \Khovanovii:
\eqn\hcoloredlong{ \ldots \longrightarrow \CH^{sl(2);V_3} (K)
\longrightarrow \CH^{sl(2)} (K^2) \longrightarrow^{\kern -11pt u}~
\Z \longrightarrow \ldots }
where $\CH^{sl(2)} (K^2)$ is the ordinary Khovanov homology of the
2-cable $K^2$ of $K$. In particular, for the 0-framed unknot we
have
\eqn\htwounknot{ \CH_{i,j}^{sl(2);V_3} ({\rm unknot})= \cases{ \Z
& if $i=0$ and $j \in \{ -2,0,2 \}$ \cr 0 & otherwise}}
This indeed agrees with the $so(3)$ homology of the unknot, if we
identify the $q$-grading in the $sl(2)$ theory with twice the
$q$-grading in the $so(3)$ theory. Therefore, combining this with
the isomorphism \sothreeisom, we expect the following formula for
the Poincare polynomial of the $so(3)$ knot homology:
\eqn\sothreeviakh{ \bar \CP^{sl(2);V_3} (K;q,t) = \bar{Kh}
(K^2;q^{1/2},t) - 1 }
We can compare this with our prediction for the $so(3)$ knot
homology based on Conjecture 1.
For example, for the trefoil knot, the specialization of the
Kauffman superpolynomial \trefoilpunred\ to $\la=q^2$ gives
\eqn\sothreetref{\eqalign{{ \bar\CF(3_1; \la= q^2,q,t) \over
(q-q^{-1})} =& q + q^2 + q^3 + q^4 t^2 + q^5 t^2 \cr &+ q^7 (t^3 +
t^4) + q^8 t^3 + q^9 t^5 + q^{10} t^5 + q^{12} t^6}}
while \sothreeviakh\ gives
\eqn\hcoloredtref{\eqalign{ \bar \CP^{sl(2);V_3} (3_1;q,t) = & q
(2 + t) + q^2 + q^3 (2 t^2 + t^3) + q^4 t^4 + q^5 (t^3 + 2 t^4) +
q^6 (t^5 + t^6) \cr & + q^7 (t^5 + t^8) + q^8 t^7 + q^9 t^9 +
q^{10} t^{11} + q^{12} t^{12} }}
It is easy to see that the structure of \sothreetref\ and
\hcoloredtref\ is very similar. However, the expression
\hcoloredtref\ based on the Khovanov homology of the 2-cable of
the trefoil contains extra terms of the form $(1+t)Q^+ (q,t)$.
It is natural to expect that in the formulation of the colored
Khovanov homology as a cohomology of the complex $\CC^{sl(2);V_3}$
these extra terms will be killed by the differential.
The remaining terms in \hcoloredtref\ agree with \sothreetref\
after a change of $t$-grading in some of the terms
(roughly speaking, in going from \hcoloredtref\ to \sothreetref,
the $t$-grading is reduced by a factor of 2, in such a way that
it does not affect specialization to $t=-1$).
The analysis of other simple knots leads to similar conclusions.

%%%%%%%%%%%%%%%%%%%%%%%%%%%%%%%%%%%%%%%%%%%%%%%%%%%%%%

\subsec{Comparison with Khovanov Homology}

Finally, let us test the relation \sptwokh, which is the
categorification of \sptwopolrel. Recall from section 2
that the $sp(2)$ quantum invariant can be obtained in two
different specializations of the Kauffman polynomial:
\eqn\twospecs{C_2(K;q) = \bar F(K;\la=-q^3,q)
= \bar F(\bar K;\la=q^{-3},q)}
Correspondingly, there are in principle two ways in which 
$sp(2)$ knot homology could be obtained from the Kauffman 
homology. We have included one of them in Conjecture 1.
Taking the trefoil as an example, one can see that the 
naive specialization to $\la=q^{-3}$ does not reproduce the
$sp(2)$ homology predicted from the isomorphism \sptwokh. This
is a first indication that $N=2$ is not ``large enough'' as
far as $sp(N)$ homology is concerned, and there are corrections 
to the naive specialization. We will see this again in section 
6 when we consider the reduced homology, for which the small 
$N$ corrections are under better control.

We cannot resist, however, to offer instead an observation
concerning the categorification of the other possible 
specialization in \twospecs, to $\la=-q^3$. For the trefoil, 
\eqn\trefspt{ {\bar\CF(3_1; \la= -q^3,q,t)\over q-q^{-1}} = 
-q^2-q^6-q^{10} t^4 + q^{18} t^6 } 
which, remarkably, is related to the ordinary Khovanov homology
in a simple way,
\eqn\trefsptt{ {\bar\CF(3_1;\la=-q^3,q,t)\over q-q^{-1}} =
- \bar{Kh} (3_1;q^2,-t^2) \,. }
Because of the sign changes, a homological interpretation of
this relation is not immediately obvious. But it is hard to 
believe that it is a pure coincidence. The relation holds 
for all knots for which we have been able to determine the 
unreduced Kauffman homology.

%%%%%%%%%%%%%%%%%%%%%%%%%%%%%%%%%%%%%%%%%%%%%%%%%%%%%%%%%%%

\newsec{Kauffman Homology}

There is also a reduced version of Conjecture 1:

\medskip\noindent
{\bf Conjecture 1$^\prime$:} {\it For a knot $K$, there exists a finite
polynomial $\SK (K) \in \Z [\la^{\pm 1}, q^{\pm 1}, t^{\pm 1}]$
such that}
\eqn\rsuperpol{\eqalign{
\HSO_N (K;q,t) &= \SK (K;\la = q^{N-1},q,t)  \cr
\HSp_N (\bar K;q,t) &= t^{s} \SK(K; \la=q^{-N-1},q,t) }}
{\it for all sufficiently large $N$.}
\foot{In the formula for the $sp(N)$ homology, $s$ denotes 
an invariant of the knot similar to Rasmussen's invariant 
\Rasmussenii. See comment below.}

In order for this latter version of the conjecture to be true, all
the coefficients of the reduced superpolynomial $\SK (\la,q,t)$
need to be non-negative. As in the $sl(N)$ case \DGR, this
suggests that $\SK(K)$ is itself a Poincare polynomial of a
triply-graded theory, whose Euler characteristic is the normalized
Kauffman polynomial. Combining this with the additional structure
of differentials inferred from the analysis of the Landau-Ginzburg
theory in section 4, we come to the following:

\medskip\noindent
{\bf Conjecture 2:} {\it There exists a triply-graded homology
theory, $\CH^{\rm Kauffman}_* =\CH^{\rm Kauffman}_{i,j,k} (K)$,
categorifying the Kauffman polynomial.
It comes with a family of differentials $\{ d_N \}$, two
further, ``universal'' differentials, $\du$ and $\duu$, and has the
following properties:}

\item{$\bullet$} categorification:
\eqn\dcat{ \chi (\CH^{\rm Kauffman}_* (K)) = F (K) }
\item{$\bullet$} anticommutativity:
\eqn\dant{ d_N d_M = - d_M d_N }
\item{$\bullet$} finite support:
\eqn\dsupp{ \dim (\CH^{\rm Kauffman}_* (K)) < \infty }
\item{$\bullet$} specializations
\eqn\dspec{
(\CH^{\rm Kauffman}_*(K),d_N) \cong
\cases{\CH^{so(N)}_* (K) \qquad & $N > 1$ \cr
\CH^{sp(-N)}_* (K) \qquad & $N < 0$ }}
\item{$\bullet$} ``universal'' and ``canceling'' differentials:
The properties we expect of $d_0$, $d_1$ and $d_2$,
as well as the two additional differentials $\du$ and $\duu$,
are explained in detail below. Roughly speaking, the cohomology of
$\CH_*^{\rm Kauffman}$ with respect to $\du$ and $\duu$ should be
isomorphic (with a simple regrading) to the HOMFLY homology
$\CH_*^{\rm HOMFLY}$,
\eqn\univtohomfly{
(\CH^{\rm Kauffman}_*(K),\duuu) \cong \CH_*^{\rm HOMFLY} }
while the cohomology with respect to $d_i$,
for $i=0,1,2$ should be ``trivial'' and depend in a particular
simple way on the knot $K$.

\noindent
Notice, that in order to be consistent with the specialization to $\la = q^{N-1}$,
the $q$-degree of the differential $d_N$ must be proportional to $(N-1)$.
in particular, this implies that the differentials $d_N$ are all trivial for sufficiently
large values of $N$, since $\CH^{\rm Kauffman}_* (K))$ has finite support.
Therefore, Conjecture 2 implies Conjecture 1$^{\prime}$,
where the superpolynomial $\CF(K;\la,q,t)$
is simply the Poincare polynomial of $\CH^{\rm Kauffman}_*$,
\eqn\gPoi{\CF(K;\la,q,t) = \sum_{i,j,k}  \la^i q^j t^k \dim\CH^{\rm Kauffman}_{i,j,k}(K)}
In terms of $\CF(K;\la,q,t)$, the condition \dcat\ can be expressed as
\eqn\expressed{\CF(K;\la,q,-1) = F(K;\la,q) }

We believe there should exist a combinatorial definition of the triply-graded
Kauffman homology with all the properties listed here. In practice, while such
a definition is not available, one can use any combination of the above axioms
as the definition, and the others as consistency checks.
This will be our approach below. Specifically,
in the rest of this section we explain in more detail various aspects
of this conjecture, use it to make predictions, and present some non-trivial checks.
We start with the discussion of the differentials.

%%%%%%%%%%%%%%%%%%%%%%%%%%%%%%%%%

\subsec{Differentials}

\noindent
{\bf $so/sp$ differentials:}

These are the differentials which justify the idea that $\CH^{\rm Kauffman}_*$
is a unified theory for $so/sp$ knot homologies. Namely, for every $N>1$
we expect the differentials $d_N$, such that  the cohomology of $\CH_*^{\rm Kauffman}$
with respect to $d_N$ yields $so(N)$ homology $\CH_*^{so(N)}$.
These differentials are expected to have degrees
\eqn\sodi{
so(N) \quad (N\ge 2): \quad {\rm deg} (d_N) = (-1,N-1,-1)  }
consistent with the specialization $\la=q^{N-1}$. Indeed, acting on the bigraded
chain complex
\eqn\sohhh{
\bigoplus_{i(N-1)+j=p} \CH_{i,j,k}^{\rm Kauffman} }
the differential $d_N$ has $q$-degree zero and $t$-degree $-1$.
The evidence for these differentials comes from the analysis of
the Landau-Ginzburg potentials in section 4, where we found that
the $so(N)$ knot homology is equipped with a family of differentials
$\{ d_{N\to M} \}$, $M<N$, which correspond to the deformations of
the Landau-Ginzburg potential,
$$
W_{so(N)} \to W_{so(N)}+x^{M-1}
$$
As we have explained in section 4, the cohomology with respect to 
$d_{N\to M}$ is expected to be isomorphic to the $so(M)$ knot homology.
These are precisely the properties of the differentials
induced on $(\CH_*^{\rm Kauffman},d_N)$ from the differentials $\{ d_M \}$
in the triply-graded theory.

Similarly, for even values of $N \le -2$ we expect the differentials,
such that the cohomology of the bigraded complex \sohhh\ with respect
to $d_N$ yields yields $sp(-N)$ homology $\CH_*^{sp(-N)}$.
As in the $sl(N)$ case \DGR, we expect the $\la$- and $q$-degree
of these differentials to be given by the same formula as for $N>1$.
Specifically,
\eqn\spdi{sp(-N) \quad (N \le -2):\quad {\rm deg} (d_N) = (-1,N-1,-1+N)}
Notice, that when the differential $d_N$ is trivial --- {\it e.g.} when its degree
is too large --- the corresponding $so(N)/sp(N)$ homology is given by \sohhh,
up to a simple re-grading. In this case, the Poincare polynomial
of the corresponding knot homology is simply a specialization
of the Kauffman superpolynomial, {\it cf.} \rsuperpol.

\ifig\trefoilfig{Dot diagram for the trefoil knot. Each dot represents a term
in the Kauffman superpolynomial; its horizontal (resp. vertical) position
encodes the power of $q$ (resp. the power of $\la$). The bottom row has $\la$-grading 2.
The differential $d_{-2}$ is represented by a solid red arrow, while the differentials
$d_0$, $d_1$, and $d_2$ are shown by dashed blue arrows. The universal differentials
$\duuu$ are depicted by curved green arrows.}
{\epsfxsize2.0in\epsfbox{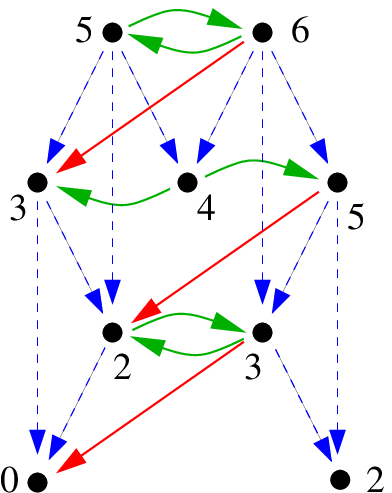}}

\example{The Trefoil Knot}
The normalized Kauffman polynomial for the trefoil knot has 9 terms:
\eqn\fnormtrefoil{ F (3_1) = \la^2 (q^2 + q^{-2}) - \la^3 (q - q^{-1})
+ \la^4 (1 - q^2 - q^{-2}) + \la^5 (q - q^{-1}) }
Similarly, using the isomorphism \sofourisoma, we find that the reduced
$so(4)$ knot homology also has rank 9, and the Poincare polynomial
\eqn\sofourredtref{
HSO_4 (3_1) = q^4 + 2 t^2 q^8 + 2 t^3 q^{10} + t^4 q^{12}
+ 2 t^5 q^{14} + t^6 q^{16}  }
Moreover, there is a unique way to identify each term
in \fnormtrefoil\ with the corresponding term in \sofourredtref,
so that their specializations to $\la=q^3$ and $t=-1$ agree.
Assuming that there are no ``hidden'' terms, we obtain the following
prediction for the reduced Kauffman superpolynomial of the trefoil knot:
\eqn\reducedcftref{
\CF (3_1) = \la^2 (q^{-2} + q^2 t^2) + \la^3 (q^{-1} t^2 + q t^3)
+ \la^4 (q^{-2} t^3 +  t^4 + q^2 t^5) + \la^5 (q^{-1} t^5 + q t^6) }
This result, based on specializations to the Kauffman polynomial
and to the $so(4)$ homological invariant, can now be used to make
predictions for other $so/sp$ homologies of the trefoil, as well as
to test the consistency of our axioms. For example, it is interesting
to note that even for such a simple knot as the trefoil,
the differential $d_{-2}$ acts non-trivially on $\CH_*^{\rm Kauffman}$,
so that the resulting homology has rank three,
\eqn\reducedsptref{ HSp_2 (\bar{3_1}) = q^4 + q^{12} t^2 + q^{16} t^3 }
in agreement with \sptwokh.
\endexample

Since $so(2)$ is abelian, the specialization to $N=2$ is expected
to give a very simple theory. In other words, it means that most
of the terms in the Kauffman homology are killed by the differential $d_2$.
A differential with this property, namely such that its cohomology
is one-dimensional, is called {\it canceling}.
In the triply-graded theory categorifying the HOMFLY polynomial \DGR,
an example of such differential is $d_1$,
which leads to the deformed theory of Lee \ESLii.
It turns out that, in the theory we are considering here, there are several
candidates for such differentials, which we discuss now.

\noindent
{\bf canceling differentials:} In a multiply-graded
theory such as the knot superhomologies, the existence of a canceling differential
of a certain degree is a rather non-trivial feature with interesting origin and
consequences. For all simple knots for which we have been able to obtain a prediction
for the triply-graded theory $\CH_*^{\rm Kauffman}$ (see below), there is room for
three canceling differentials, which can be naturally identified as $d_2$, $d_1$ and $d_0$,
and which have degrees,
\eqn\dcgrad{\eqalign{
d_2&: \qquad (-1,1,-1)  \cr
d_1&: \qquad (-2,0,-3) \cr
d_0&: \qquad (-1,-1,-2)}}
Moreover, the one-dimensional piece surviving the differentials sits in a particular
degree, which depends in a simple way on the knot $K$. Namely, we find that, for
all the knots that we considered, the reduced superpolynomial for the Kauffman
homology can be written in three different ways
\eqn\forms{\eqalign{
\SK (\la,q,t) &= (\la/q)^{-s} + (1+\la q^{-1}t) Q_2^+ (\la,q,t) \cr
\SK (\la,q,t) &= (\la t)^{-2s} + (1+\la^2 t^3 ) Q_1^+ (\la,q,t) \cr
\SK (\la,q,t) &= (\la q t)^{-s} + (1+\la q t^2) Q_0^+ (\la,q,t)
}}
where the polynomials $Q_i^+$ ($i=0,1,2$) have integer non-negative coefficients.
As an example, we show how the canceling differentials
act on the Kauffman homology for the trefoil knot in \trefoilfig.

It is interesting to note that the triple-grading of the term 
surviving any given canceling differential is related in a simple 
way to the degree of that differential. Namely,
\eqn\simpledegree{
\deg\bigl[ (\CH^{\rm Kauffman},d_i)\bigr] = -s[\deg(d_i) - (0,0,1)]
\qquad {\rm for} \; i=0,1,2}
A similar relation holds for the canceling differentials of the
HOMFLY theory \DGR.

Based on our examples and on the experience with the $sl(N)$ case \DGR,
it is natural to conjecture that the Kauffman homology of any knot should
admit three canceling differentials. A more conservative form of the conjecture
would allow the cohomology with respect to $d_i$'s, $i=0,1,2$, to be not strictly
one-dimensional, or only so after addition of a non-homogeneous correction.

\noindent
{\bf Universal differentials:}
The existence of the universal differentials is perhaps the most novel
and intriguing aspect of Conjecture 2. Indeed, equipped with these
differentials, the Kauffman homology should not only be a unified
framework for $so/sp$ homologies, but in fact should also contain
the triply graded HOMFLY theory!

The evidence for at least one universal differential comes from
the ``universal'' deformation of the Landau-Ginzburg potential
discussed in section 4:
$$
W_{so(N)} \to W_{so(N)} + y^2
$$
This deformation suggests the existence of a differential $d_{y^2}$
that takes $so(N)$ homology to $sl(N-2)$ homology. Moreover,
since this differential should exist for all values of $N$,
it is natural to expect that it is induced from a differential
in the triply graded theory that relates $\CH^{\rm Kauffman}_*$
and $\CH^{\rm HOMFLY}_*$, so that the differentials relating
$so(N)$ and $sl(N-2)$ homology are just different specializations
of this universal differential.

Surprisingly, the Kauffman homology appears to have two universal
differentials. Namely, for all knots that we considered,
the Kauffman superpolynomial can be related to the HOMFLY
superpolynomial in two different ways:
\eqn\univ{\eqalign{
\SK(\la,q,t) &= q^{-s} \CP(\la/q,q,t) + (1+q^2 t) \Quu^+ (\la,q,t) \cr
\SK(\la,q,t) &= (qt)^s \CP(\la q t,q,t) + (1+q^{-2} t^{-1}) \Qu^+(\la,q,t)}}
where $\Quu^+$ and $\Qu^+$ are polynomials with integer non-negative coefficients.
These two relations suggest two universal differentials, that we denote 
$\duu$ and $\du$, respectively, with gradings
\eqn\dugrad{
{\rm deg}(\duu) = (0,-2,-1) \qquad
 {\rm deg}(\du) = (0,2,1) }
It is curious to note that by specializing \univ\ to $t=-1$,
we obtain a non-trivial relation between the Kauffman and HOMFLY polynomial,
which to the best of our knowledge is new.

Finally, we should point out that the numbers $s$ in eqs. \rsuperpol, \forms,
and \univ\ might be different knots invariants. However, as in the $sl(N)$
case \DGR, we conjecture that they are all equal and use the same notation $s(K)$.
Furthermore, in all the examples that we considered, this invariant
is actually equal to the knot signature, $s(K) = \sigma (K)$,
which is also familiar from the structure of the $sl(N)$ homological invariants.

%%%%%%%%%%%%%%%%%%%%%%%%%%%%%%%%%%%%%%%%%%%%%%%%%%%%%%%%%%%%%%%%%%%%%%%%%%

\subsec{Thin Knots and $\delta$-grading}

Taking some of the axioms \dcat\ - \univtohomfly\ as a definition of
the triply-graded homology $\CH_*^{\rm Kauffman}$, we can predict what
it should be for a number of simple knots, as we did for the trefoil
in \reducedcftref. For example, in the following table below we write
the Kauffman superpolynomial for all knots with less than seven crossings.
These results can be used to test other properties of the Kauffman
homology, in particular, the structure implied by the canceling,
universal, and $sp(2)$ differentials.

These predictions reveal another property of the Kauffman homology,
which it shares with other theories.
Namely, all the existing knot homologies have an interesting property
(which, in a sense, is a hint for unification) that the structure of
a theory often becomes simpler if instead of the ordinary homological
grading ($t$-grading) one introduces a new grading --- usually called
$\delta$-grading --- which is a linear combination of the original gradings.
Roughly speaking, the $\delta$-grading tells us about the homological
complexity of a knot. For example, different homological invariants
of a knot with small number of crossings are all localized in one
value of the corresponding $\delta$-grading.
Such knots are called {\it homologically thin}, or {\it thin} for short.
The first example of a knot which is not thin\foot{Sometimes,
such knots are called {\it thick}.} is the 8-crossing knot $8_{19}$.

The $\delta$-grading in the Kauffman homology is a linear combination
of $\la$, $q$ and $t$ gradings,
\eqn\deltadef{\delta = {3\over 2}\la + {1\over 2}q - t }
In particular, it is easy to verify that all the knots listed
in Table 3 are thin in Kauffman homology, and their $\delta$-grading 
coincides with minus the signature of the knot.
The first example of a thick knot in Kauffman homology
is again the knot $8_{19}$.

%%%%%%%%%%%%%%%%%%   TABLE STARTS HERE   %%%%%%%%%%%%%%%%%%%%%%%
\vfill \eject
%%%%%%%%%%%%%%%%%%%%%%%%%%%%%%%%%%%%%%%%%%%%%%%%%%%%%%%%%%%%%%%%

\vskip 0.8cm \vbox{ \centerline{\vbox{
\hbox{\vbox{\offinterlineskip
%%% usual height for tablespace is 2pt
\def\tablespace{height7pt&\omit&&\omit&&\omit&\cr}
\def\tablerule{\tablespace\noalign{\hrule}\tablespace}

\hrule\halign{&\vrule#&\strut\hskip0.2cm\hfill
#\hfill\hskip0.2cm\cr \tablespace & Knot   && ~~$s$~ && $\CF (\la,q,t)$ &\cr
\tablerule
& $3_1$  &&  $-2$  && $\la^2 (q^{-2} + q^2 t^2) + \la^3 (q^{-1} t^2 + q t^3)
+ \la^4 (q^{-2} t^3 +  t^4 + q^2 t^5) + \la^5 (q^{-1} t^5 + q t^6)$ &\cr \tablerule
& $4_1$   &&  $0$  && $\la^{-2} (q^{-2} t^{-4}+ t^{-3} + q^2
t^{-2}) + \la^{-1} (q^3 + q^{-3} t^{-3} + 2 q^{-1} t^{-2} + 2 q
t^{-1})$  & \cr &   &&  && $+ 2 q^{-2} t^{-1} + 3 +  2 q^2 t + \la
(q^{-3} + 2 q^{-1} t + 2 q t^2 + q^3 t^3) + \la^2 (q^{-2} t^2 +
t^3 + q^2 t^4)$ &\cr \tablerule
& $5_1$  &&  $-4$   && $\la^4 (q^{-4} + t^2 + q^4 t^4) + \la^5 (q^{-3} t^2 +
q^{-1} t^3 + q t^4 + q^3 t^5)$ & \cr &   &&  && $+ \la^6 (q^{-4} t^3 +
q^{-2} t^4 + 2 t^5 + q^2 t^6 + q^4 t^7) + \la^7 (q^{-3} t^5 + 2
q^{-1} t^6 + 2 q t^7 + q^3 t^8)$ &  \cr &   &&   && $+ \la^8 (q^{-2} t^7
+ 2 t^8 + q^2 t^9) + \la^9 (q^{-1} t^9 + q t^{10})$ &\cr
\tablerule
& $5_2$  &&  $-2$  && $\la^2 (q^{-2} + t + q^2 t^2) + \la^3 (q^{-3} t + 3
q^{-1} t^2 + 3 q t^3 + q^3 t^4)$ & \cr &   &&  && $+ \la^4 (q^{-4} t^2 +
3 q^{-2} t^3 + 5 t^4 + 3 q^2 t^5 + q^4 t^6) + \la^5 (2 q^{-3} t^4
+ 4 q^{-1} t^5 + 4 q t^6 + 2 q^3 t^7)$ & \cr &   &&  && $+ \la^6 (q^{-4}
t^5 + 2 q^{-2} t^6 + 3 t^7 + 2 q^2 t^8 + q^4 t^9) + \la^7 (q^{-3}
t^7 + q^{-1} t^8 + q t^9 + q^3 t^{10})$ &\cr \tablerule
& $6_1$  &&  $0$  && $\la^{-2} (q^{-2} t^{-4} + t^{-3} + q^2 t^{-2}) +
\la^{-1} (q^3 + q^{-3} t^{-3} + 3 q^{-1} t^{-2} + 3q t^{-1}) +
q^{-4} t^{-2}$ & \cr &   &&  && $+ 4 q^{-2} t^{-1} + 6 + 4 q^2 t + q^4
t^2 + \la (3 q^{-3} +  q^{-5} t^{-1} + 6 q^{-1} t + 6 q t^2 + 3
q^3 t^3 + q^5 t^4)$ & \cr &   &&  && $+ \la^2 (2 q^{-4} t + 4 q^{-2} t +
5 t^3 + 4 q^2 t^4 + 2 q^4 t^5)$ & \cr &   &&  && $+ \la^3 (q^{-5} t^2 +
2 q^{-3} t^3 + 3 q^{-1} t^4 + 3 q t^5 + 2 q^3 t^6 + q^5 t^7)$ &
\cr &   &&  && $+ \la^4 (q^{-4} t^4 + q^{-2} t^5 + t^6 + q^2 t^7 + q^4
t^8)$ &\cr \tablerule
& $6_2$  &&  $-2$  && $q^4 + q^{-4} t^{-4} + q^{-2} t^{-3} + 2 t^{-2} + q^2
t^{-1}$ & \cr &   &&  && $+ \la (4 q + q^{-5} t^{-3} + 3 q^{-3} t^{-2} +
4 q^{-1} t^{-1} + 3 q^3 t + q^5 t^2)$ & \cr &   &&  && $+ \la^2 (6
q^{-2} + 3 q^{-4} t^{-1} + 8 t + 6 q^2 t^2 + 3 q^4 t^3)$ & \cr &
  &&  && $+ \la^3 (q^{-5} + 5 q^{-3} t + 9 q^{-1} t^2 + 9 q t^3 + 5 q^3
t^4 + q^5 t^5)$ & \cr &   &&  && $+ \la^4 (2 q^{-4} t^2 + 6 q^{-2} t^3 +
9 t^4 + 6 q^2 t^5 + 2 q^4 t^6)$ & \cr &   &&  && $+ \la^5 (2 q^{-3} t^4
+ 5 q^{-1} t^5 + 5 q t^6 + 2 q^3 t^7) + \la^6 (q^{-2} t^6 + 2 t^7
+ q^2 t^8)$ &\cr \tablerule
& $6_3$  &&  $0$  && $\la^{-3} ( q^{-3} t^{-6} + 2 q^{-1} t^{-5} + 2 q
t^{-4} + q^3 t^{-3} )$ & \cr &   &&  && $+ \la^{-2} (2 q^{-4} t^{-5} + 5
q^{-2} t^{-4} + 7 t^{-3}  + 5 q^2 t^{-2} + 2 q^4 t^{-1} )$ & \cr &
  &&  && $+  \la^{-1}  (6 q^3 + q^{-5} t^{-4} + 6 q^{-3} t^{-3} + 11
q^{-1} t^{-2} + 11 q t^{-1} + q^5 t )$ & \cr &   &&  && $+ 4 q^{-4}
t^{-2} + 10 q^{-2} t^{-1} + 15 + 10 q^2 t + 4 q^4 t^2$ & \cr &   &&  &&
$+ \la ( 6 q^{-3} + q^{-5} t^{-1} + 11 q^{-1} t  + 11 q t^2 + 6
q^3 t^3 + q^5 t^4)$ & \cr &   &&  && $+ \la^2 (2 q^{-4} t + 5 q^{-2} t^2
+ 7 t^3 + 5 q^2 t^4 + 2 q^4 t^5)$ & \cr &   &&  && $+ \la^3 (q^{-3} t^3
+ 2 q^{-1} t^4 + 2 q t^5 + q^3 t^6)$ &\cr
 }\hrule}}}}
\centerline{ \hbox{{\bf Table 3:}{\it ~~ Kauffman superpolynomial
for some simple knots.}}} } \vskip 0.5cm

%%%%%%%%%%%%%%%%%%%%%%%%%%%%%%%%%%%%%%%%%%%%%%%%%%%%%%%%%%%%%%%%
\vfill \eject
%%%%%%%%%%%%%%%%    TABLE ENDS HERE   %%%%%%%%%%%%%%%%%%%%%%%%%%

Further examples of thin knots are torus knots of type $(2,2k+1)$.
Their reduced Kauffman superpolynomial is given by the following
general formula:
\eqn\fredtorusknots{
\CF (T_{2k+1,2}) = (\la t)^{4k} + (\la^2 t^3 + 1) \Big( {\la \over q}
\Big)^{2k} \Big[ \sum_{i=0}^k q^{4i} t^{2i} + \sum_{j=1}^{2k-1}
\sum_{i=1}^{2k-j+1} \la^j q^{j+2i-2} t^{2j+i-1} \Big]
}
It is easy to verify that the structure of this result
is consistent with all the specializations
and the differentials that we proposed.
Also, it is curious to note that, at least for these torus knots,
there is a relation between the reduced and unreduced Kauffman
superpolynomials, which is similar to the relation between
$\CP (K)$ and $\bar \CP (K)$ implied by the existence of
the canceling differential $d_1$ \DGR.
Specifically, we can write \fredtorusknots\ in the following form:
\eqn\form{
\CF (K) = (\lambda/q)^{2k} + (1+\lambda q^{-1} t) \Bigl[
(1+\lambda q^{-3}) Q_a + (1+q^2 t) Q_b \Bigr]}
As in the HOMFLY case, the structure of this expression also
reflects the existence of the canceling differential $d_2$
in the Kauffman homology, but it is strictly stronger
than \forms\ and is written in terms of {\it two} polynomials:
\eqn\qqtorus{\eqalign{
Q_a &= \lambda^{2k} \sum_{i=1}^k t^{2i} q^{4i-2k} \cr
Q_b &= \la^{2k+2} \sum_{j=0}^{k-1}\;\sum_{i=0}^{2k-2j-2} t^{2i+2j+4}\la^i q^{i+4j-2k+2}
%Q_b &=\lambda^{2k+2} \sum_{i=0}^{2k-2} t^{2i+4}\lambda^iq^{i-2k+2}
%\sum_{j=0}^{{\rm min}(i,2k-i-2)} \lambda^{-j}q^{3j}
}}
Notice, that $Q_a$ also appears in the expression for the HOMFLY
superpolynomial of the same torus knots \DGR. 
The unreduced Kauffman superpolynomial \funredtorusknots\
for these torus knots can be also written in terms of $Q_a$ and 
$Q_b$:
\eqn\sform{
\eqalign{\bar\CF (K) = &(\lambda-\lambda^{-1} + q-q^{-1})(\lambda/q)^{2k}
+ (1+\lambda q t) \Bigl[(1+\lambda q^{-3}) (1-q^2 \lambda^{-2}) \lambda q^{-2} Q_a \cr
& + (1+\lambda^2 q^{-2} t)(1-q^{-2}) \lambda^{-1}q^2 Q_b\Bigr]
}}
Unfortunately, this relation between $\CF (K)$ and $\bar \CF (K)$
does not extend to other thin knots. For example, even though
the reduced Kauffman superpolynomial for the figure-eight knot
has the form analogous to \form, the corresponding expression does 
not agree with \figeight\ (in fact, it does not even reduce to the 
$so(4)$ knot homology at $\la=q^3$).
It would be interesting to study the relation between the reduced
and unreduced Kauffman superpolynomials further.

As another application of the $\delta$-grading, let us derive
the regrading in the relation \univ\ between the Kauffman and
HOMFLY superpolynomials. The $\delta$-grading of a thin knot
in the HOMFLY theory is $\delta =\la+{1\over 2}q-t=-s/2$,
while the $\delta$-grading of the same knot in the Kauffman
theory is $\delta={3\over 2}\la+{1\over 2}q -t=-s$.
Since $q$ and $t$ have $\delta$-grading ${1\over 2}$ and $-1$
in both theories, the factors $q^{-s}$ and $(qt)^s$ shift
the $\delta$-grading by $-s/2$, as required for consistency.
Similarly, since in the Kauffman theory $\la$ has $\delta$-grading
${3 \over 2}$, the combinations $\la/q$ and $\la q t$ have
$\delta$-grading ${1 \over 2}$, which is precisely what
the $\delta$-grading of $\la$ should be in the HOMFLY theory.

Finally, notice that the differentials $d_N$ and $d_{-N}$, $|N|>1$,
have the same $\delta$-grading.
Moreover, note that the differential $d_{-2}$, the universal differentials $\duuu$,
and the canceling differentials $d_0$, $d_1$, and $d_2$ all have zero $\delta$-grading,
which means that they can be non-trivial even for thin knots
(see {\it e.g.} \trefoilfig).

%%%%%%%%%%%%%%%%%%%%%%%%%%%%%%%%%%%%%%%%%%%%%%%%%%%%%%%%%%%%%%%%%%%%%%%

\vskip 30pt

\centerline{\bf Acknowledgments}

\noindent
We would like to thank M.~Khovanov, M.~Mari\~no, C.~Vafa, and 
E.~Witten for valuable discussions.
We are grateful to the KITP, Santa Barbara for warm 
hospitality during the program ``Mathematical Structures in String 
Theory'', where part of this work was carried out.
This work was conducted during the period
S.G. served as a Clay Mathematics Institute Long-Term Prize Fellow.
This work was also supported in part by the DOE under grant
number DE-FG02-90ER40542, in part by RFBR grant 04-02-16880
and in part by the NSF under Grant No. PHY99-07949.

\listrefs
\end